\documentclass[extra]{gji}
\usepackage{color,graphicx,mathrsfs,times,url,lineno}
\usepackage[fleqn]{amsmath}
\usepackage{amssymb}
\newcommand{\unboxed}{}
\newcommand{\also}{\qquad\mbox{and}\qquad}
\newcommand{\alss}{\quad\mbox{and}\quad}
\newcommand{\for}{\qquad\mbox{for}\qquad}
\newcommand{\where}{\quad\mbox{where}\quad}
\newcommand{\with}{\quad\mbox{with}\quad}

\newcommand{\dbk}{\,d\kb}
\newcommand{\dkb}{\,d\kb}
\newcommand{\dkkp}{\delta({\kb,\kb'})}
\newcommand{\dkp}{\,dk'}
\newcommand{\dk}{\,dk}
\newcommand{\dtbk}{\dkb\dkb'}
\newcommand{\dxb}{\,d\xb}

\newcommand{\hsomm}{\hspace{-0.1em}}
\newcommand{\hsom}{\hspace{0.1em}}

\newcommand{\hspst}{\hspace{0.25em}}
\newcommand{\hsps}{\hspace{0.75em}}
\newcommand{\hth}{\hat{\theta}}
\newcommand{\htwo}{0.3775\textheight}
\newcommand{\wtwo}{0.725\textwidth}

\newcommand{\intnyq}{\int\!\!\!\!\int}
\newcommand{\intnyqk}{\iint\limits_\kb}
\newcommand{\intok}{\int_0^{k}}
\newcommand{\ka}{k_\alpha}
\newcommand{\kbp}{\mathbf{k'}}
\newcommand{\kb}{\mathbf{k}}
\newcommand{\mcC}{{\mathcal{C}}} 
\newcommand{\mcF}{{\mathcal{F}}}
\newcommand{\mcH}{{\mathcal{H}}} 
\newcommand{\mcJ}{\mathcal{J}}
\newcommand{\mcL}{{\mathcal{L}}}
\newcommand{\mcN}{{\mathcal{N}}}
\newcommand{\mcP}{{\mathcal{P}}}
\newcommand{\mcS}{{\mathcal{S}}} 
\newcommand{\nff}{\left(\frac{\Dx\Dy}{MN}\right)^\half}
\newcommand{\nffs}{\left(\frac{\Dx\Dy}{MN}\right)}
\newcommand{\nfi}{\frac{2\pi}{(MN\Dx\Dy)^{\half}}}
\newcommand{\norml}{\frac{1}{MN}\sum_{\kb}}
\newcommand{\normltu}{\frac{1}{(MN)^2}\sum_{\kb}\sum_{\kbp}}

\newcommand{\ofarguzh}{(z/2)}
\newcommand{\ofarguz}{(z)}
\newcommand{\ofargu}{\left(\argu\right)}
\newcommand{\ofkp}{(\kbp)}
\newcommand{\ofk}{(\kb)}

\newcommand{\ofskp}{(k')}
\newcommand{\ofsk}{(k)}
\newcommand{\ofsr}{(r)}
\newcommand{\ofst}{_{\btheta}}

\newcommand{\ofsy}{(y)}
\newcommand{\ofth}{(\hbt)}
\newcommand{\oftr}{(\btruth)}

\newcommand{\oft}{(\btheta)}
\newcommand{\ofx }{(\xb)}
\newcommand{\ofxp}{(\xb')}

\newcommand{\ofz}{(0)}
\newcommand{\bF}{\mathbf{F}}

\newcommand{\bxp}{\mathbf{x^{\prime}}}
\newcommand{\bx}{\mathbf{x}}
\newcommand{\by}{\mathbf{y}}
\newcommand{\bzero}{\mathbf{0}}

\newcommand{\sQ}{\mathsf{Q}}
\newcommand{\xbp}{\mathbf{x'}}
\newcommand{\xb}{\mathbf{x}}
\newcommand{\yb}{\mathbf{y}}

\newcommand{\bbF}{{\mbox{\boldmath$\bar{\mathcal{F}}$}}}
\newcommand{\bbg}{{\mbox{\boldmath$\bar{\gamma}$}}}
\newcommand{\bgamma}{{\mbox{\boldmath$\gamma$}}}
\newcommand{\bmcF}{\mbox{\boldmath$\mathcal{F}$}}
\newcommand{\bmcJ}{\mbox{\boldmath$\mathcal{J}$}}
\newcommand{\bnabla}{\mbox{\boldmath$\nabla$}}

\newcommand{\btheta}{\boldsymbol{\theta}}
\newcommand{\Dx}{\Delta x}
\newcommand{\Dy}{\Delta y}
\newcommand{\Tit}{^{\it{\sst{T}}}}
\newcommand{\arguni}{\big(2n^{1/2}r\big)\left/\left(\pi\rho\right)\right.}

\newcommand{\argu}{\frac{2\nu^\half}{\pi\rho}r}
\newcommand{\plth}{\pl_\theta}
\newcommand{\pl}{\partial}
\newcommand{\shalf}{^{1/2}}
\newcommand{\sst}{\scriptstyle}
\newcommand{\stx}{s^2_X}
\newcommand{\st}{{\sigma^2}}
\newcommand{\tpi}{\frac{1}{2\pi}}

\newcommand{\tpti}{\frac{1}{(2\pi)^2}}
\newcommand{\tpt}{(2\pi)^2}

\newcommand{\truth}{\theta_0}
\newcommand{\Jthth }{\mcJ_{\theta\theta}}
\newcommand{\Sbar}{\bar\mcS}
\newcommand{\bgthp}{\bar{\gamma}_{\theta'}}
\newcommand{\bgth}{\bar{\gamma}_{\theta}}

\newcommand{\bmthp}{\bar{m}_{\theta'}}
\newcommand{\bmth}{\bar{m}_{\theta}}
\newcommand{\bthetaSfull}{[\st\,\,\nu\,\,\,\rho\,]\Tit}
\newcommand{\btruth}{\btheta_0}
\newcommand{\cov}{\mbox{cov}}

\newcommand{\gth}{\gamma_{\theta}}
\newcommand{\half}{\frac{1}{2}}

\newcommand{\hbt}{{\mbox{\boldmath$\hat{\theta}$}}}
\newcommand{\inlaw}{\stackrel{\mathrm{law}}{\longrightarrow}}
\newcommand{\mthp}{m_{\theta'}}
\newcommand{\mth}{m_{\theta}}

\newcommand{\var}{\mbox{var}}
\newcommand{\edc}{\end{document}}
\onecolumn
\title[Statistical analysis of geophysical fields]
{Maximum-likelihood estimation of the Mat\'ern covariance
  structure of isotropic spatial random fields on finite, sampled grids}
\author[F.~J.~Simons et al.]{
  \begin{minipage}[]{\textwidth}{Frederik J.~Simons$^{1,2}$, Olivia
      L.~Walbert$^1$, Arthur P.~Guillaumin$^3$, Gabriel L.~Eggers$^4$, Kevin
      W.~Lewis$^5$, and Sofia C.~Olhede$^6$\\[-0.5em]}
  \end{minipage}\\  
  $^1$ Department of Geosciences, Princeton University, Princeton, NJ 08544,  USA. E-mail: fjsimons@alum.mit.edu\\
  $^2$ Program in Applied \& Computational Mathematics, Princeton University, Princeton, NJ 08544, USA\\
  $^3$ Queen Mary University of London, E1 4NS, London, UK\\
  $^4$ Wesleyan University, Middletown, CT 06459, USA\\
  $^5$ Earth and Planetary Sciences, Johns Hopkins University, Baltimore, MD 21218, USA\\
  $^6$ Ecole Polytechnique F\'ed\'erale de Lausanne, Switzerland}

\begin{document}
\maketitle

\begin{summary}
We present a statistically and computationally efficient spectral-domain
maximum-likelihood procedure to solve for the structure of Gaussian spatial
random fields within the Mat\'ern covariance hyperclass. For univariate,
stationary, and isotropic fields, the three controlling parameters are the
process variance, smoothness, and range. The debiased Whittle likelihood
maximization explicitly treats discretization and edge effects for finite
sampled regions in parameter estimation and uncertainty quantification. As even
the `best' parameter estimate may not be `good enough', we provide a test for
whether the model specification itself warrants rejection. Our results are
practical and relevant for the study of a variety of geophysical fields, and for
spatial interpolation, out-of-sample extension, kriging, machine learning, and
feature detection of geological data. We present procedural details and
high-level results on real-world examples.
\end{summary}


\section{I~n~t~r~o~d~u~c~t~i~o~n}
\label{introduction}

What numbers, which statistical notions capture the ``essence'' of a spatial
patch of geophysical data? If it were a stationary, white, Gaussian process,
simply, its population mean and variance would be sufficient. However, computed
over differing window sizes, or at varying resolution, sample means and
variances of geophysical data sets fluctuate non-erratically, hence
``whiteness'' immediately proves to be an untenable assumption. Reporting
summary statistics over changing baselines \cite[e.g.,][for Earth and planetary
  topography]{Sharpton+85,Aharonson+98,Grohmann+2010,Rosenburg+2011} effectively
subscribes to the data as realizations from a spatially \textit{correlated}
(locally stationary, Gaussian) process. Estimating even just the variance of
said field (that is, at a point), with little bias (that is, accurately) and
with a reasonable estimation variance (that is, precisely), requires knowledge
of the co-variance (between pairs of points), and an acknowledgment of the
inherently finite nature of the available spatial patch over which a single
(geological) process can be thought of as representative. We are led to the
estimation of (the parameters of) a spatial covariance function, or
alternatively and equivalently, of a spectral density function, from (hopefully
locally) stationary sampled data that are (irregularly) bounded \cite[see,
  e.g.,][]{Valentine+2020}. The first (spatial estimation) is notoriously noisy
since it requires finding data pairs at increasing offsets, and computationally
time-consuming when it entails inverting a covariance matrix \cite[see,
  e.g.,][]{Mardia+84,Kitanidis+85,Vecchia88,Grainger+2021}. The second (spectral
estimation), while typically faster and less noisy, is famously affected by
aliasing, finite-field, and edge effects
\cite[e.g.,][]{Stein95,Appourchaux+98a,Appourchaux+98b,Hamilton2009a,Hamilton2009b}.
Fourier-domain artifacts lead to estimation bias especially for multidimensional
data sets, where the nefarious influence of boundary terms dominates the
mean-squared error \cite[]{Dahlhaus+87}.

Estimating model parameters of random fields typically relies on
spatial~\cite[]{Kent+96} or spectral~\cite[]{Whittle53} likelihood
maximization. If the former is often slow to compute, and the latter suffers
from potential boundary effects, there are notable modifications to both. For
the spatial methods, Vecchia approximations to Gaussian log-likelihoods are fast
and accurate \cite[see, e.g.,][]{Katzfuss+2021,Porcu+2024}.  Circulant-embedding
based methodologies~\cite[]{Guinness+2017,Guinness2019} effectively recompute
the Fourier basis rather than assuming it known from homogeneity in
space. Vecchia-type approaches do require a local neighborhood for prediction,
chosen appropriately. Some of these methods are discussed relative to spectral
approaches both in theory and via simulations by~\cite{Guillaumin+2022}, who
furthermore introduce adjustments to correct for partial-grid and boundary
effects with the spectral methods, developing a large-sample theory for the
appropriate asymptotic framework, upon which we build in this paper.

Common choices for parameterized covariance functions of Gaussian random fields
are exponential or squared-exponential forms
\cite[e.g.,][]{Tarantola+84,Montagner86,Gudmundsson+90,Baig+2003,Baig+2004a},
defined solely by a variance and a correlation range. Both are special cases of
the \cite{Matern60} class of covariance functions \cite[]{Guttorp+2006}, with a
specific smoothness or mean-squared differentiability
\cite[]{Adler1981,Christakos92}, itself an inversion parameter of interest in a
multitude of geophysical application domains such as seismology
\cite[e.g.,][]{Wu+85b,Wu+90,Becker+2007b,Carpentier+2007}, seafloor bathymetry
and oceanography \cite[]{Goff+89a,Goff+89b,Goff+2010,Sandwell+2022,Simon+2026},
helioseismology \cite[e.g.,][]{Gizon+2004}, hydrology
\cite[e.g.,][]{Rodriguez+74,Kitanidis+85}, meteorology, and climate science
\cite[e.g.,][]{Handcock+94,Paciorek+2006,Lindgren+2011,North+2011,Sun+2015}.

In this paper we present a univariate spectral-domain debiased ``Whittle''
maximum-likelihood procedure \cite[]{Simons+2013,Guillaumin+2017,Sykulski+2019}
that estimates the variance, smoothness, and range of an isotropic Mat\'ern
Gaussian process, from sampled spatial data. We show how to obtain unbiased
estimates for these size, scale, and shape parameters when the region studied is
neither rectangular nor circular, nor completely sampled. We exactly calculate
their estimation covariance, correctly blurred for finite-field effects and
without neglecting commonly omitted wave vector correlation effects, such that
the results from differently sized and sampled patches can be compared
robustly. We develop and comprehensively illustrate our algorithms for
completely sampled rectangular data sets.

While process variance estimation necessitates recognizing the effects of
spatial covariance, which is aided by the ability to extend the size of the
observation domain, intuitively ``beyond the reach of the range'', conversely,
to characterize the smoothness of a process requires neighboring observations
that are highly correlated. In other words, both increasing (growing) domain and
fixed domain (infill) asymptotics \cite[and mixed-domain, see][]{Chu2023} are
relevant considerations \cite[see, e.g.,][]{Zhang2004,Zhang+2005b} that we are,
however, unable to fully address in this paper, although the computer code that
we freely distribute (see Sec.~\ref{code}) is designed to give the reader all
the tools in hand to be able to do so. Neither form of asymptotics is a panacea,
and, furthermore, an unfortunate aspect of the Mat\'ern class is that in two
dimensions, neither the variance nor the range are consistently estimable under
fixed-domain asymptotics, although there is a non-linear function of them and
the smoothness that is \cite[see][]{Stein95}. \cite{Guillaumin+2022}, who
provided the theoretical basis for our work, introduced the notion of
\textit{significant correlation contribution} as a means of capturing whether a
covariance function under a given spatial sampling mechanism allows us to
distinguish parameter values. It cannot eliminate intrinsic problems with the
flexibility of the Mat\'ern family, and in some scenarios it may indeed be
beneficial to \textit{fix} the smoothness. While the full theoretical framework
allows us to determine when we are able to estimate a specific Mat\'ern family
with a given sampling mechanism, it is our goal, with this paper, to provide
sufficient, but not exhaustive, guidance, accompanied by computer code, to make
Mat\'ern covariance estimation practical and sensible for geoscientists.

Our results have widespread implications for the study of geophysical fields,
and should be interpreted in the light of our trying to derive ``process'' from
``parameters'', the end goal being to be able to assign likely formation
mechanisms and histories for the patches under consideration. Our results should
also be relevant for whomever needs to perform spatial interpolation or
out-of-sample extension (e.g., via kriging) on geological data
\cite[]{Journel+78,Christakos92,Cressie93,Stein99}. They also carry consequences
for machine learning and feature detection
\cite[]{Rasmussen+2006,Porcu+2024}. We present procedural details but also focus
on high-level results that have real-world applications. We illustrate our
methodology on four geologically and geophysically relevant data sets, assuming
stationarity within patches that have been selected via user interpretation.

\section{P~R~E~L~I~M~I~N~A~R~I~E~S}
\label{basics}

Readers wishing to come to terms with the geological, geophysical, and geodetic
definitions of `relief', `topography', or `elevation' are directed to
\cite{Lambeck88}, \cite{Hofmann+2006}, and \cite{Wieczorek2015}. To make the
jump from geology and geophysics to statistics, particularly in this context, we
first and foremost recommend (re)reading \cite{Goff+88,Goff+89a}, who also
discuss anisotropic processes. The material in this section is both an extension
and a specialization of the multivariate results of \cite{Simons+2013}, which is
to be consulted for further details. Here, we use a more explicit notation,
adapt some of the normalizations, and make a number of modifications---but most
importantly, we restrict our analysis to univariate two-dimensional Cartesian
isotropic Gaussian fields.

\subsection{Continuous framework}

Here we draw most heavily on Sections~2.1 and~4.1 of \cite{Simons+2013}.
Referring furthermore to \cite{Percival+93}, \cite{Stein99}, and
\cite{Vanmarcke2010} for additional considerations and terminology, to
\cite{Abramowitz+65} and \cite{Gradshteyn+2000} for properties of special
functions, we begin by defining the particular quantities of interest in the
\textit{spatial} and the \textit{spectral} domains.

\subsubsection{Stationarity}
\label{station}

A geophysical field $\mcH\ofx$ is considered to be a zero-mean, finite-variance,
stationary, two-dimensional random field with finite second moments. Under what
is known as the \cite{Cramer42} representation, there exists a spectral
orthogonal-increment process, $d\mcH\ofk$, according to which the spatial field
\begin{equation}\label{cramer}
\mcH\ofx =
\intnyq
e^{i\kb\cdot\xb}\,d\mcH\ofk,\qquad \xb\in\mathbb{R}^2.
\end{equation} 
The integration is over the space containing all wave vectors~$\kb$. In the case
of strict bandlimitation or very fast spectral decay we may restrict the
computations to the Nyquist plane $[-\pi,\pi]\times[-\pi,\pi]$. The expectation
of $d\mcH\ofk$ over many realized fields,
\begin{equation}
\label{zeromean}
\langle d\mcH\ofk\rangle=0,
\end{equation}
and its variance, in the absence of covariance between wave vectors, defines a
power-spectral density, $\mcS\ofk$, in the form of the expectation
\begin{equation}
\label{specdens}
\langle d\mcH^{}\ofk \hsom d\mcH^*\!\ofkp\rangle=
\mcS\ofk\dtbk\,\dkkp
,
\end{equation}
where $\dkkp$ is the Dirac delta function. When
eqs~(\ref{cramer})--(\ref{specdens}) hold, the spatial auto-covariance,
$\mcC(\xb,\xbp)$, displays stationarity by being dependent on
\textit{separation}, $\xb-\xbp$, only, since in that case we can write for the
expectation of the two-point spatial-domain products
\begin{equation}\label{covsep} 
\langle \mcH\ofx \mcH^*\hsomm\ofxp\rangle=\intnyq e^{i\kb\cdot(\xb-\xbp)}\mcS\ofk\dbk=\mcC(\xb-\xbp)
.  
\end{equation}

The spectral variance (at the wave vectors~$\kb$) and the spatial covariance (in
the \textit{lag} variables~$\xb$) form a Wiener-Khintchine Fourier pair,
\begin{align}
\label{wiener1}\mcC(\xb)&=\intnyq e^{ i\kb\cdot\xb}  S(\kb)\dkb,\\
\label{wiener2}\mcS(\kb)&=\tpti\intnyq \mcC(\xb) \,e^{-i\kb\cdot\xb}\dxb.
\end{align}
The zero-wavenumber intercept of the spectral density is the zeroth
moment of the spatial covariance:
\begin{equation}\label{zerom} 
\mcS\ofz=\tpti\intnyq \mcC(\xb)\dxb .
\end{equation}

\subsubsection{Isotropy}

Under isotropy, abusing notation, $\mcS\ofk=\mcS\ofsk$, depending only on the
scalar wavenumber $k=\|\kb\|$. Integrating over the polar angles to bring out
$J_0$, the Bessel function of the first kind and of order zero, the spatial
covariance,
\begin{equation}\label{Ciso}
\langle \mcH\ofx \mcH^*\hsomm\ofxp\rangle
=2\pi\int J_0(k \|\xb-\xbp\|)\, \mcS\ofsk \, k\dk
=\mcC(\|\xb-\xbp\|),
\end{equation}
is dependent only on \textit{distance}, $\|\xb-\xbp\|$, not
direction. Since $J_0(0)=1$, the isotropic spatial variance is then
given by
\begin{equation}\label{variclass}
\langle \mcH\ofx \mcH^*\hsomm\ofx\rangle=
2\pi\int \mcS\ofsk \,k\dk
=\mcC(0)=\st
.  
\end{equation}
Introducing the distance variable~$r$ and integrating over the angular polar
coordinate, we rewrite eq.~(\ref{zerom}) as
\begin{equation}\label{zeromi} 
\mcS\ofz=\tpi\int \mcC\ofsr\,r\,dr.
\end{equation}
We follow \cite{Vanmarcke2010} in adopting the term `fluctuation
scale' for $\mcS(0)/\st$, the mass of the spatial
\textit{correlation} function $\mcC(r)/\st$.

Isotropy remains a restrictive---but testable---assumption, which we will be
relaxing in forthcoming work. Any future discussion of anisotropy will entail
evaluating it against the null-hypothesis of isotropic behavior, in the possible
presence of anisotropic sampling patterns, for which the present paper provides
all the necessary statistical machinery.

\subsubsection{Mat\'ernity}

We further specify the field as a member of the Mat\'ern class \cite[]{Stein99},
which is very general and widely applicable to processes with monotonically
decreasing autocovariance \cite[]{Guttorp+2006}. The isotropic $d$-dimensional
Mat\'ern spectral density~$\mcS^d\ofst\ofsk$ assumes the parameterized form
\cite[]{Handcock+94,WangKesen+2023}
\begin{equation}\label{maternd} 
\mcS^d\ofst\ofsk=\frac{\Gamma(\nu+d/2)}{\Gamma(\nu)}
\frac{\st}{\pi^{d/2}}\left(\frac{4\nu}{\pi^2\rho^2}\right)^{\nu}
\left(\frac{4\nu}{\pi^2\rho^2}+k^2\right)^{-\nu-d/2}
,
\end{equation}
where $\Gamma$ is the gamma function, and which, in two dimensions, $d=2$, as we
subsumed earlier and maintain from now on, specifies to
\begin{equation}\label{matern2} 
\mcS\ofst\ofsk=
\st\frac{\pi\rho^2}{4}\left(\frac{4\nu}{\pi^2\rho^2}\right)^{\nu+1}
\left(\frac{4\nu}{\pi^2\rho^2}+k^2\right)^{-\nu-1}.
\end{equation}
With this model, our principal unknowns are its three strictly
positive parameters, denoted generically as $\theta>0$, which we
collect in the set
\begin{equation}\label{thetaS}
\btheta=\bthetaSfull
.
\end{equation}
The `variance', $\st$, indeed satisfies eq.~(\ref{variclass}) upon substitution
with eq.~(\ref{matern2}). At short wavelengths, when $k$ is large, the spectrum
$\mcS\ofst\ofsk$ decays at a rate that depends on the `smoothness', $\nu$, which
expresses the $\lceil \nu-1\rceil$ times (mean-squared) `differentiability' of
the process \cite[]{Handcock+93}. The behavior at the longest wavelengths, for
small $k$, is controlled by the combined effect of $\st$ and~$\rho$. The
fluctuation scale
\begin{equation}\label{matern0} 
\frac{\mcS\ofst(0)}{\st}=\frac{\pi\rho^2}{4}
.
\end{equation}

The isotropic Mat\'ern spatial covariance $\mcC\ofst\ofsr$, in terms of the lag
distance~$r$, is unlike its spectral counterpart~(\ref{maternd}) in requiring no
dimensional specification, that is, independently of $d$,
\begin{equation}\label{Kr}
\mcC\ofst\ofsr=\st\frac{2^{1-\nu}}{\Gamma(\nu)}\ofargu^{\nu}K_\nu\ofargu
,
\end{equation}
with $K_\nu$ the modified Bessel function of the second kind. The asymptotic
behavior $K_\nu(z)\rightarrow\Gamma(\nu)\,(z/2)^{-\nu}/2$ for small~$z$,
verifies that $\mcC\ofst(0)=\st$ as in eq.~(\ref{variclass}). For low values of
$\nu$, furthermore, $\mcC\ofst(\pi\rho)\approx\st/3$. In other words, spatial
correlations generally die down by a factor of about two-thirds at distances
beyond $r\approx\pi\rho$, hence the name for the third parameter, the
`correlation length' or `range', $\rho$.

The power accumulated over a certain wavenumber interval, counting from the
origin, is given by the distribution function
\begin{equation}\label{power}
\mcP\ofst\ofsk 
=
2\pi\intok \mcS\ofst\ofskp \,k' \dkp
=\st\left[1-\left(\frac{4\nu}{\pi^2\rho^2}\right)^\nu\left(\frac{4\nu}{\pi^2\rho^2}+k^2\right)^{-\nu}\right]
.
\end{equation}
As expected $\mcP\ofst(0)=0$ and $\mcP\ofst(\infty)=\st$. We define the
wavenumbers $\ka$ at which the power reaches $100\times\alpha$ per
cent of the total,
\begin{equation}\label{getit}
\mcP\ofst(\ka)=\alpha\st
,
\end{equation} 
which, from eq.~(\ref{power}), is solved analytically by
\begin{equation}\label{kasolve} 
\ka=\frac{2\nu^\half}{\pi\rho}\left[(1-\alpha)^{-1/\nu}-1\right]^\half
.
\end{equation}
It can be readily verified that $k_{0}=0$ and $k_{1}=\infty$. For convenience,
we express the equivalent wavelengths in the notation
$\lambda_{100\alpha}=2\pi/\ka$.

The flexible generality of the Mat\'ern class is appreciated by evaluating the
correlation functions for special values of~$\nu$
\cite[]{Guttorp+2006}. Notably, when $\nu=1/2$, the correlation function decays
exponentially, and when $\nu\rightarrow\infty$, as a Gaussian---a squared
exponential. Other examples include the Von K\'arm\'an ($\nu=1/3$), Whittle
($\nu=1$), and second-order ($\nu=3/2$) and third-order ($\nu=5/2$)
autoregressive correlation models. Despite all of its generality and wide
applicability, the exponentially decaying, isotropic, three-parameter Mat\'ern
model cannot capture all random fields under study. It is not identifiable with
(multi-) fractal, scale-invariant, self-affine, or self-similar behavior
\cite[see, e.g.,][]{Mareschal89,Herzfeld+95,Gneiting+2012,Landais+2019}, as it
is self-affine only at high wavenumber. For temporal processes,
\cite{Lilly+2017} discuss when the Mat\'ern process is more appropriate to use
than fractional Brownian motion. Developing a general set of methods outside
these constraints, e.g., when the spectrum is no longer
bounded~\cite[]{Reed+2002}, remains outstanding as future work.

\begin{figure*}\centering
  \includegraphics[width=0.9\textwidth,angle=0]{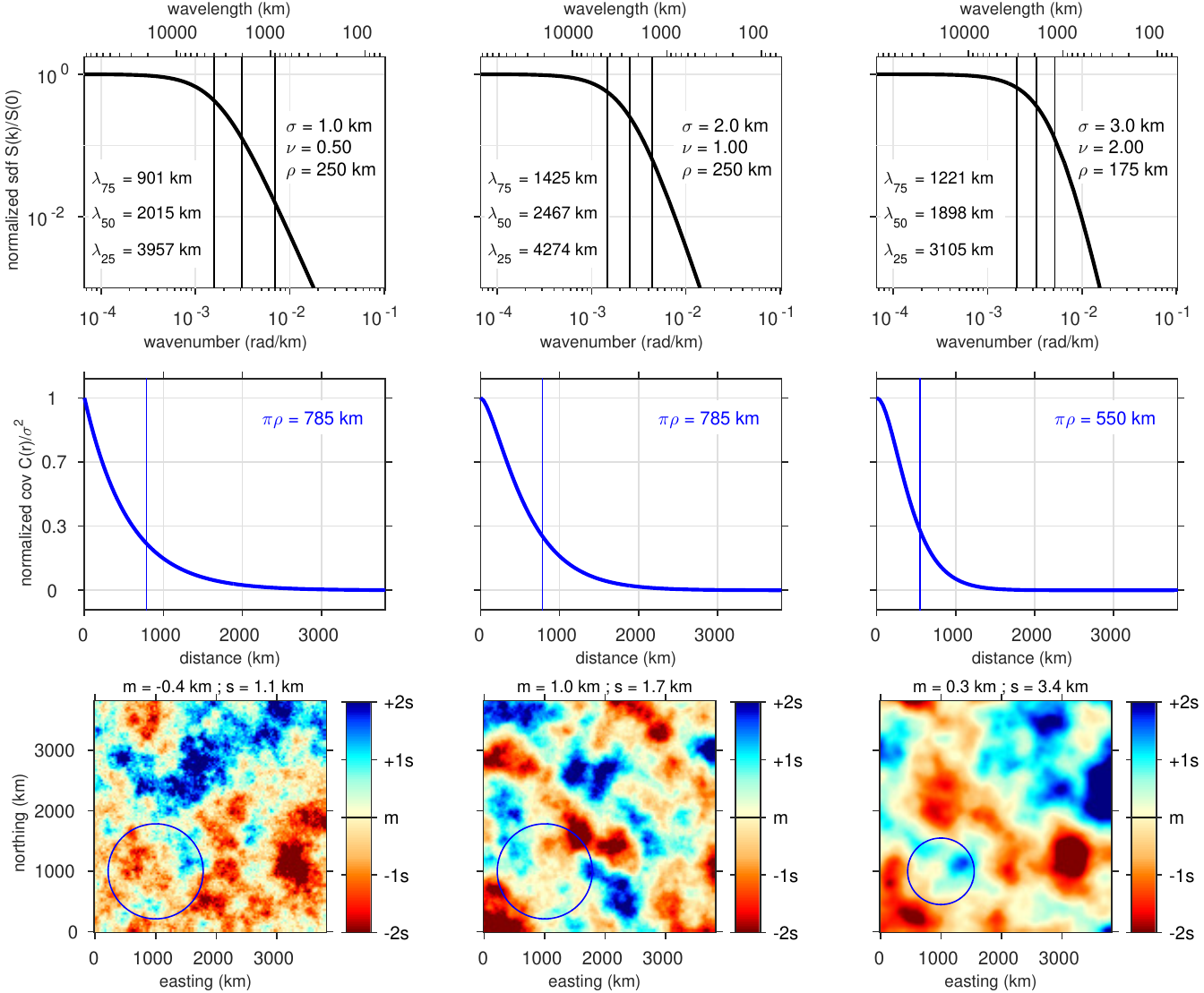}
  \caption{\label{sdfcovetc}Random fields generated from stationary isotropic
    Mat\'ern models with variances $\st$, differentiabilities $\nu$, and
    correlation lengths~$\rho$, as indicated.  (\textit{Top:}) Normalized
    spectral densities, $\mcS\ofst\ofsk/\mcS\ofst(0)$, from eq.~(\ref{matern2}).
    The vertical black lines identify the wavenumbers $\ka$ at which the power
    reaches $100\times\alpha$ per cent of the variance, from
    eq.~(\ref{kasolve}), in wavelengths $\lambda_{100\alpha}=2\pi/\ka$, as
    labeled. (\textit{Middle:}) Correlations, the normalized spatial
    covariances, $\mcC\ofst(r)/\st$, from eq.~(\ref{Kr}). The vertical blue
    lines are drawn at the values $\pi\rho$, the distances at which the
    correlations die down to approximately one third of the
    variance. (\textit{Bottom:}) Field realizations.  The blue circles have
    radii $\pi\rho$, drawn for visual guidance. In the titles, $\mathsf{m}$ and
    $\mathsf{s}$ identify the sample means and standard deviations.
}
\end{figure*}

Fig.~\ref{sdfcovetc} provides intuitive insight into the role that the three
parameters $\st$, $\nu$, and $\rho$ play in the spatial behavior of Mat\'ern
random fields, synthesized by the circulant embedding procedure outlined in the
next section.

\subsection{Lattice framework}

Here we rely mostly on sections~2.1, 4.2, and A6 of \cite{Simons+2013}.  The
properties of the finite and sampled, i.e., \textit{windowed} discrete
processes, as will be experienced in computational data analysis, differ
markedly from the behavior of the idealized, infinite, continuous models
discussed in the previous section, and those two viewpoints need to be
explicitly reconciled.

\subsubsection{Discretization}

For simplicity of notation, $\xb$ now maps out a rectangular $K=M\times N$ grid
of ``pixels'' with spatial extent $\Dx$ and $\Dy$ given by
\begin{equation}\label{xgrid} 
\xb=\big\{
(m\,\Dx, n\,\Dy)
\big\},
\for 
m=0,\dots,M-1
\also 
n=0,\dots,N-1.
\end{equation}
We define the discrete Fourier transform of the noiseless measurements of the
spatial process $\mcH\ofx$ obtained after sampling as
\begin{equation}\label{fourierH} 
H\ofk
\equiv
\frac{1}{2\pi}\nff\sum_\xb \mcH\ofx e^{-i\kb\cdot\xb}
.  
\end{equation}
Sampled in spectral space, the finite set of wave vectors is now, with $m$ and $n$
as in eq.~(\ref{xgrid}),
\begin{equation}\label{kgrid}
\kb=
\big(k_x,k_y\big)=
\left\{
\left(
\frac{2\pi}{M\Dx}
\left[-\left\lfloor\frac{M}{2}\right\rfloor+m\right],\,
\frac{2\pi}{N\Dy} 
\left[-\left\lfloor\frac{N}{2}\right\rfloor+n\right]
\right)
\right\}
,
\end{equation}
and on this complete Nyquist grid we now identify eq.~(\ref{cramer}),
consistently with eq.~(\ref{fourierH}), with
\begin{equation}\label{cramers} 
\mcH\ofx
\equiv
\nfi\sum_\kb e^{i\kb\cdot\xb}\hsom H\ofk
.% No more R2
\end{equation} 

\subsubsection{Blurring}\label{blurring}

Obtaining space-domain realizations from a population of random fields specified
by a certain spectral density such as given by eq.~(\ref{matern2}) is possible
by generating Fourier coefficients~$H\ofk$, as in eq.~(\ref{fourierH}), directly
on the spectral grid~(\ref{kgrid}), and by inverse Fourier transformation, as in
eq.~(\ref{cramers}), onto the spatial grid~(\ref{xgrid}). These $H\ofk$ should
be drawn from a zero-mean complex proper Gaussian distribution, with expectation
zero, $\langle H\ofk\rangle=0$, and with a covariance $\langle H\ofk
H^*\hsomm\ofk\rangle$ that will be influenced by the chosen size, shape and
discretization of the region under consideration; i.e., it will be different
from the theoretical quantity $\langle d\mcH^{}\ofk\hsom
d\mcH^*\hsomm\ofkp\rangle$ of eq.~(\ref{specdens}), which involved the true
density~$\mcS\ofst\ofk$.

\cite{Simons+2013} showed the covariance of a finite set of gridded Fourier
coefficients can at best offer a blurred and correlated version of the true
spectral variance \cite[see also][]{Fournier+2014}. To us, their eq.~(9) reads,
using $\mcS\ofst\ofsk=\mcS\ofst\ofk$ from our eq.~(\ref{matern2}),
\begin{equation}\label{specblur}
  \langle H\ofk H^*\hsomm\ofkp\rangle=
  \intnyqk D_K(\kb-\kb'')D_K^*(\kb'-\kb'')\hsom\mcS\ofst(\kb'')\dbk''
  \with
  D_K\ofk=\frac{1}{2\pi}\nff\sum_\xb e^{-i\kb\cdot\xb}
  ,
\end{equation}
whereby~$D_K(\kb)$ is the Dirichlet kernel, the suitably normalized discrete
Fourier transform of the unit sampling operator. They simulated fields by
incorporating the blurring but ignoring the correlation, that is, following
their eqs~(12) and~(83), they approximated eq.~(\ref{specblur}) as
\begin{equation}\label{specblur2}
\langle H\ofk H^*\hsomm\ofkp\rangle\approx
\delta(\kb,\kb')\intnyqk \left|D_K(\kb-\kb')\right|^2\mcS\ofst\ofkp \dbk'
=\Sbar\ofst\ofk\delta(\kb,\kb')
.
\end{equation}
The blurred spectral density~$\Sbar\ofst\ofk$ is the convolution of the true
spectral density~$\mcS\ofst\ofk$ with the Fej\'er kernel
$\left|D_K(\kb-\kb')\right|^2$, which embodies the effects both of sampling and
finite domain size, and reduces to the Dirac delta for continuous processes on
infinite domains. \cite{Simons+2013} carried out the blurring
\textit{approximately}, via grid refinement, convolution, and subsampling.

\cite{Guillaumin+2022}, in contrast, in their Lemmata~1 and~2, showed how to
\textit{exactly} incorporate the spectral blurring effect of applying arbitrary
data windows \cite[see also][]{Fuentes2007}, including irregular boundaries and
incomplete sampling, at a much reduced computational cost. We rewrite the
discrete Fourier transform in eq.~(\ref{fourierH}), which silently selected
samples on the grid~$\xb$ but did not modify or weight them, to incorporate a
non-constant, unit-normalized, data window, $w\ofx$, as follows:
\begin{equation}\label{fourierHw} 
H\ofk
\equiv
\frac{1}{2\pi}\nff\sum_\xb w\ofx\mcH\ofx e^{-i\kb\cdot\xb}
. 
\end{equation}
Using the definition in eq.~(\ref{covsep}), the sample variance of the windowed
Fourier coefficients is
\begin{align}\label{varHk1}
\var\left\{H\ofk\right\}
&=\frac{1}{(2\pi)^2}\nffs
\sum_\xb\sum_\xbp w(\xb)w(\xbp) \hsom\mcC\ofst\left(\xb-\xbp\right) e^{-i \kb \cdot (\xb-\xbp)},\\
&\label{varHk2}=\frac{1}{(2\pi)^2}\nffs
\sum_\yb \left(\hspst\sum_{\xb-\yb\,\cap\,\xb} w(\xb)w(\xb-\yb)\right) \hsom\mcC\ofst\left(\yb\right) e^{-i \kb \cdot \yb},
\end{align}
following a change of variables and a change in the order of
summation, noting that the first sum is over the separation grid
\begin{equation}\label{ygrid} 
\yb=\big\{
(m'\Dx, n'\Dy)
\big\},
\qquad\mbox{with the mirrored index sets}\qquad
m'=-M+1,\dots,M-1 
\alss
n'=-N+1,\dots,N-1,
\end{equation}
and the second sum, for each element of $\by$, over the subset of
$\bx-\by$ that belongs to the original grid $\bx$, so as to stay
within the original integration domain. Our manipulations allow us to
isolate and sum out the interior term, which we rewrite more explicitly as
\begin{equation}\label{wyb}
W(\yb)
=\hspst\sum_{\max(\xb-\yb,|\yb|)} w(\xb)w(\xb-|\yb|)
=\hspst\sum_{\min(\xb-\yb,|\yb|)} w(\xb)w(\xb+|\yb|).
\end{equation}
Reviewing what this implies for the variance of the Fourier
coefficients of a windowed and sampled field $\mcH(\xb)$, we rewrite
eqs~(\ref{varHk1})--(\ref{varHk2}) as
\begin{align}\label{varHk3}
\var\left\{H\ofk\right\}
&=\frac{1}{(2\pi)^2}\nffs
\sum_\yb W(\yb) \hsom\mcC\ofst\left(\yb\right) e^{-i \kb \cdot \yb}
=\Sbar\ofst\ofk.
\end{align}
Eq.~(\ref{varHk3}) is the exact version, valid for arbitrary data windows, of
what \cite{Simons+2013} implemented approximately and only for rectangular
windows by discrete convolution of the theoretical spectral density
$\mcS\ofst\ofk$ with the Fej\'er kernel $\left|D_K(\kb)\right|^2$, as in
eq.~(\ref{specblur2}). In that special case of a unitary window function, that
is, for complete observations on a rectangular grid, eq.~(\ref{wyb}) evaluates
to the triangular
\begin{equation}
W(\yb)
=(M-|m'|)(N-|n'|).
\end{equation}

Generating Fourier coefficients~$H\ofk$ from the square root of the
\textit{blurred} spectral density~$\Sbar\ofst\ofk$ of eq.~(\ref{varHk3}), as
\cite{Simons+2013} did as a basis for simulation, ignores wavenumber
\textit{correlation} effects. In that case samples need to be generated on a
spatial grid larger than needed, retaining only a central portion for analysis,
to avoid wrap-around correlations or periodic realizations. Constructing spatial
patches via eq.~(\ref{cramers}) on the space grid~(\ref{xgrid}), the covariance
of the results, $\langle \mcH\ofx\mcH\hsomm\ofxp\rangle$, may be understood as a
discrete approximation of the integral in eq.~(\ref{covsep}), with the spacings
defined in eq.~(\ref{kgrid}). Hermitian symmetry guarantees that the simulated
fields are real, and their covariance
\begin{equation}\label{covreal}
  \langle\mcH\ofx\mcH\hsomm\ofxp\rangle=\mcC\ofst(\xb-\xbp)
  ,
\end{equation}
is stationary. Furthermore, for large sample sets,
$\sum_\kb\Sbar\ofst\ofk\approx\sum_\kb\mcS\ofst\ofk$, which establishes the
desired correspondence
\begin{align}\label{varHx}
\var\{\mcH\ofx\}\approx
\frac{\tpt\sum_\kb \mcS\ofst\ofk}{(MN\Dx\Dy)}
\approx
\st
.
\end{align}

\begin{figure*}\centering
\includegraphics[width=0.65\textwidth]{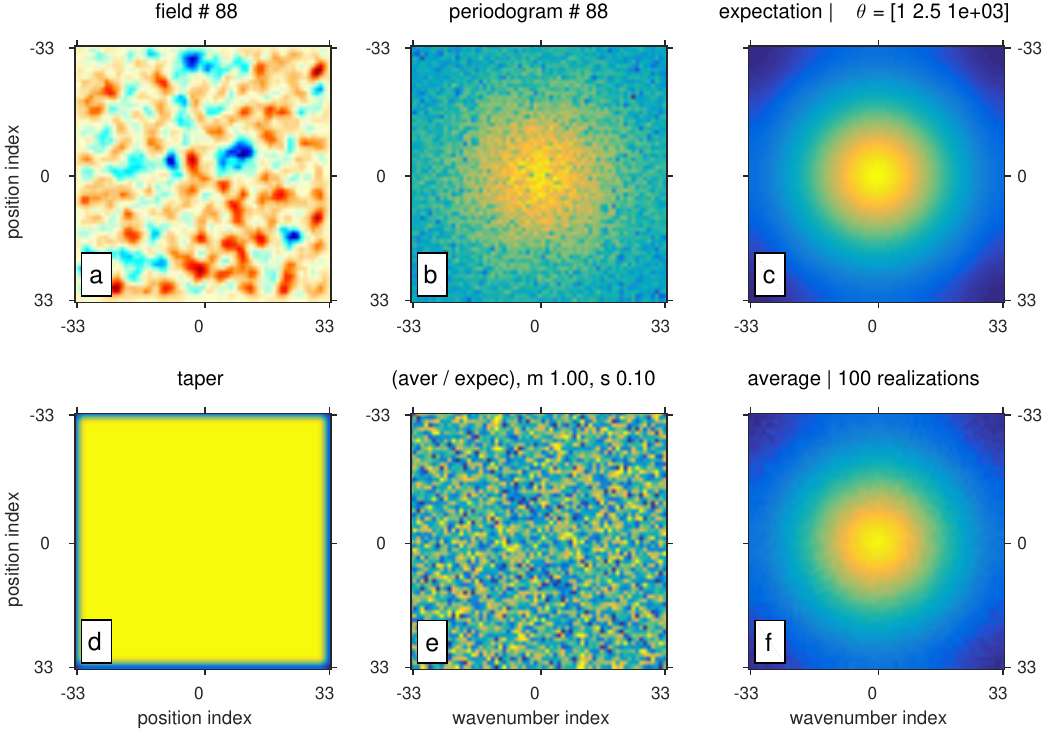}
\caption{\label{square}The Fej\'er blurred spectral density $\Sbar\ofst\ofk$
  approximates the expectation of the periodogram $|H\ofk|^2$, of gridded and
  cosine-tapered data generated from the population density
  $\mcS\ofst\ofk$. (\textit{a})~A single realization $\mcH\ofx$,
  (\textit{b})~its modified periodogram $|H\ofk|^2$, and (\textit{c}) the blurred
  spectrum $\Sbar\ofst\ofk$. (\textit{d})~The cosine-tapered unit window,
  (\textit{e})~the ratio of the average periodogram to the blurred spectral
  density, and~(\textit{f}) the average periodogram, over 100 realizations.}
\end{figure*}

Eq.~(\ref{varHk3}) shows that the \textit{expected periodogram} of the data,
$\var\{H\ofk\}=\langle|H\ofk|^2\rangle$, can be obtained via Fourier
transformation of the autocovariance sequence of the sampling
window. Fig.~\ref{square} shows this equivalence. From a sequence of
simulations, we show, in Fig.~\ref{square}\textit{(a--c)}, one spatial-domain
field $\mcH\ofx$, its periodogram $|H\ofk|^2$ on the corresponding normalized
Fourier grid, and the expected periodogram, the blurred spectrum
$\Sbar\ofst\ofk$, for the parameter set $\btheta$ shown at the top, $\st=1$ (in
arbitrary field units), $\nu=2.5$ and $\rho=1$ (in units of the spatial grid
spacing). In Fig.~\ref{square}\textit{(d--f)}, we show a unit-normalized square
window cosine-tapered around the edges, the ratio of the average periodogram to
its expectation, $\mathrm{mean}\{|H\ofk|^2\}/\Sbar\ofst\ofk$, with its sample
mean $\mathsf{m}$ and standard deviation $\mathsf{s}$, and the average of the
periodograms across 100 realizations, the sample variance $\var\{H\ofk\}$, which
approximates the blurred spectrum $\Sbar\ofst\ofk$ shown directly above.

\subsubsection{Simulation}\label{correlation}

Eq.~(\ref{specblur}) shows that the covariance of a finite set of gridded
Fourier coefficients suffers both from blurring by the sampling kernel, as we
have just illustrated and calculated explicitly, but also from correlation
between the wavenumbers. To prepare for what is coming, we note, first, that if
the off-diagonal terms in $\langle H\ofk H^*\hsomm\ofkp\rangle$ decay rapidly
enough, one could ignore them as the basis for simulations, taking $\langle
H\ofk H^*\hsomm\ofk\rangle\approx\Sbar\ofst\ofk$ as a point of departure,
whether calculated on the interior domain of a doubled grid, using grid
refinement, discrete convolution, and subsampling to approximate $\Sbar\ofst$
\cite[as was done by][]{Simons+2013}, or exactly, via eq.~(\ref{varHk3}). Here
we do take wavenumber correlations into account for data simulation by switching
to space-domain methods that use the \textit{spatial} covariance,
eq.~(\ref{Kr}), as their point of departure, via circulant embedding of the
covariance matrix \cite[]{Kroese+2015}.  Second, we will show empirically that
we are able to ignore them when designing the debiased-Whittle likelihood
\cite[]{Guillaumin+2017,Sykulski+2019} to perform parameter estimates from
sampled data, which is a central feature of this paper and its
predecessors. Finally, we show that they will, however, play an important role
in the calculation of the estimation variance of the maximum-likelihood
estimates, using the results obtained by \cite{Guillaumin+2022}, and we discuss
various algorithms to conduct the relevant calculations. This last fact stands
in apparent contradiction to the material discussed by \cite{Simons+2013}, their
Sections~A6 and A8, which, in retrospect, have proven to be overly optimistic,
asymptotically.

\section{W~H~I~T~H~E~R{\hsps}W~H~I~T~T~L~E{\hsps}?}

In this paper we develop a maximum-likelihood procedure that takes gridded input
`topographies' and estimates the three-element sets~$\btheta$, see
eq.~(\ref{thetaS}), that contain the parameters of the isotropic Mat\'ern
spectral densities $\mcS\ofst\ofsk$ or spatial covariances $\mcC\ofst\ofsr$ by
which we aim to sufficiently describe such planetary data patches. Before
proceeding, we take a brief detour to illustrate, for the example of the
variance, $\st$, \textit{why} we embark on this journey. Additional motivation
and considerations are offered by, among others, \cite{Vanmarcke83} and
\cite{Stein99}.

The variance $\st$ of a demeaned sample patch is \textit{not} well estimated by
its area-averaged sum of squares, which would amount to
\begin{equation}
\label{svar}
s^2=\frac{1}{MN}\sum_\xb \mcH^2\ofx-\frac{1}{(MN)^2}\bigg(\sum_\xb \mcH\ofx\bigg)^2
.
\end{equation}
Indeed, the expectation of the `sample variance' estimator, $s^2$, is
biased by the co-variance between the samples, which itself is
unknown. Using eqs~(\ref{covreal}) and~(\ref{varHx}), we find from
eq.~(\ref{svar}) that in expectation, approximately,
\begin{align}
\langle s^2\rangle
&\approx\frac{1}{MN}\sum_\xb\frac{\tpt\sum_\kb \mcS\ofst\ofk}{(MN\Dx\Dy)}
-\frac{1}{(MN)^2}\sum_\xb\sum_\xbp \mcC\ofst(\xb-\xbp)
%% \label{svar1}
\approx\st-\frac{1}{(MN)^2}\sum_\xb\sum_\xbp \mcC\ofst(\xb-\xbp) 
\label{svar2}\\
&\approx\st-\frac{\tpt\Sbar\ofst(0)}{MN\Dx\Dy}
.
\label{svar3}
\end{align}

\begin{figure*}\centering
\includegraphics[width=\wtwo,angle=-0]{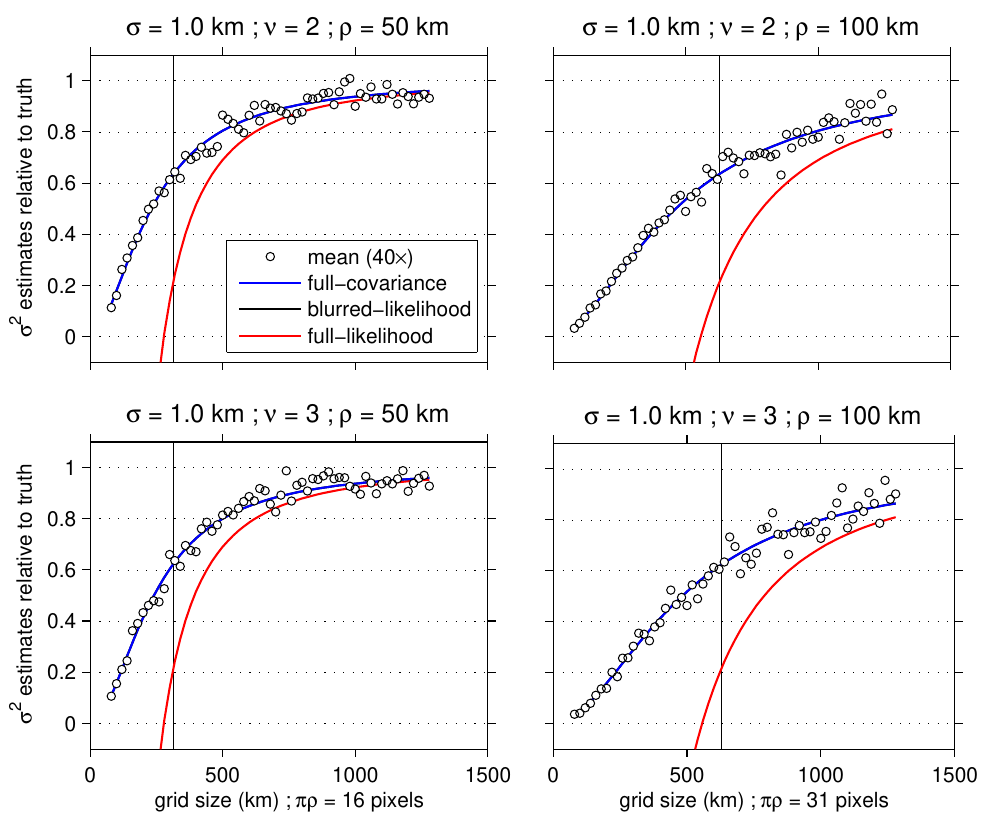}
\caption{\label{varbias}The sample variance $\mathrm{s^2}$ systematically
  underestimates the true process variance $\st$. It is negatively biased by the
  presence of spatial correlation embodied by the Mat\'ern parameters $\nu$ and
  $\rho$, listed in the titles. Black open circles (`mean') are averages of the
  sample variances $\mathrm{s^2}$ for data patches of different sizes, as
  observed over 40 lattice simulations, normalized by the actual
  variance~$\st$. Solid blue lines (`full-covariance') predict the average
  behavior by incorporating the bias according to eq.~(\ref{svar2}), evaluating
  eq.~(\ref{Kr}) on all of the pairwise distances available in the
  grids. Essentially hidden underneath the blue ones are solid black lines
  (`blurred-likelihood') resulting from calculations that use
  eq.~(\ref{svar3}). Solid red lines (`full-likelihood') are from
  eq.~(\ref{svar6}). As detailed in the text, the quality of the various
  approximations is to be interpreted in terms of the Mat\'ern correlation
  parameters $\nu$ and $\rho$, in relation to the sampling spacings $(\Dx,\Dy)$,
  which were kept constant at 10~km, and the field sizes $(M,N)$, which
  increased from left to right, as shown. The vertical black lines are drawn at
  the values $2\pi\rho$, a distance beyond which the bias in the sample variance
  estimator decreases to about a third of the true value, speaking empirically.}
\end{figure*}

Eq.~(\ref{svar2}) applies quite generally, to stationary processes with spectral
density~$\mcS\ofst\ofk$ or covariance function~$\mcC\ofst\ofx$. The error made
in reducing the last term in~(\ref{svar2}) to the second term in
eq.~(\ref{svar3}) should be interpreted as arising from the discretization of
eq.~(\ref{zerom}). The appearance of the \textit{blurred} spectral density
$\Sbar\ofst(0)$ is due to the finite-sample effects by which the spatial grid is
\textit{relatively} coarse, and too small to comprise the lags at which the
structure is completely decorrelated. Only for, \textit{effectively},
uncorrelated white noise, $\mcC\ofst(\xb-\xbp)=\st\delta(\xb,\xbp)$, does
eq.~(\ref{svar2}) reduce to the independent and identically distributed bias
expression \cite[]{Bendat+2000}
\begin{equation}\label{svar5}
\langle s^2\rangle\approx
\st\left(1-\frac{1}{MN}\right)
.
\end{equation}
If the spatial grid is fine enough, that is, if the pixel sizes $\Dx$ and $\Dy$
are small enough relative to $\rho$, the scale length of the correlation, thus
when the full behavior of the spatial covariance $\mcC\ofst\ofx$ is being
accurately captured by the sampling density, $\mcS\ofst(0)$ can again be
substituted for $\Sbar\ofst(0)$ in eq.~(\ref{svar3}). In that case, using
eq.~(\ref{matern0}) yields the form applicable to the isotropic Mat\'ern
density, namely
\begin{align}\label{svar6}
\langle s^2\rangle
&\approx\st\left(1-\frac{\pi(\pi\rho)^2}{MN\Dx\Dy}\right)
.
\end{align}
While this last approximation is usually too crude for bias calculations,
eq.~(\ref{svar6}) does explain the expected behavior that, the larger $\rho$,
relative to the area of the study region, the more correlation will be present
between the samples, causing us to underestimate the variance, and the more
negatively biased the naive estimator eq.~(\ref{svar}) will be.  In real-world
applications we will of course know neither the variance~$\st$ nor the
range~$\rho$. Nor the smoothness~$\nu$, for that matter, knowledge of which
might otherwise help us design better estimators, with $\nu$ held fixed.

Fig.~\ref{varbias} illustrates the arguments made so far in this section, for a
variety of values of $\nu$ and $\rho$, as a function of grid size, and where the
expectation of the estimate is formed by averaging over a great number of
experiments. The naive variance estimator~$s^2$ is biased, in a manner and for a
reason that we understand intuitively, and are able to compute
analytically. Rather than writing down expressions for the \textit{variance} of
the naive variance estimator~$s^2$, we will illustrate its behavior on the basis
of another suite of numerical experiments.

\begin{figure*}\centering
\includegraphics[width=\wtwo,angle=-0]{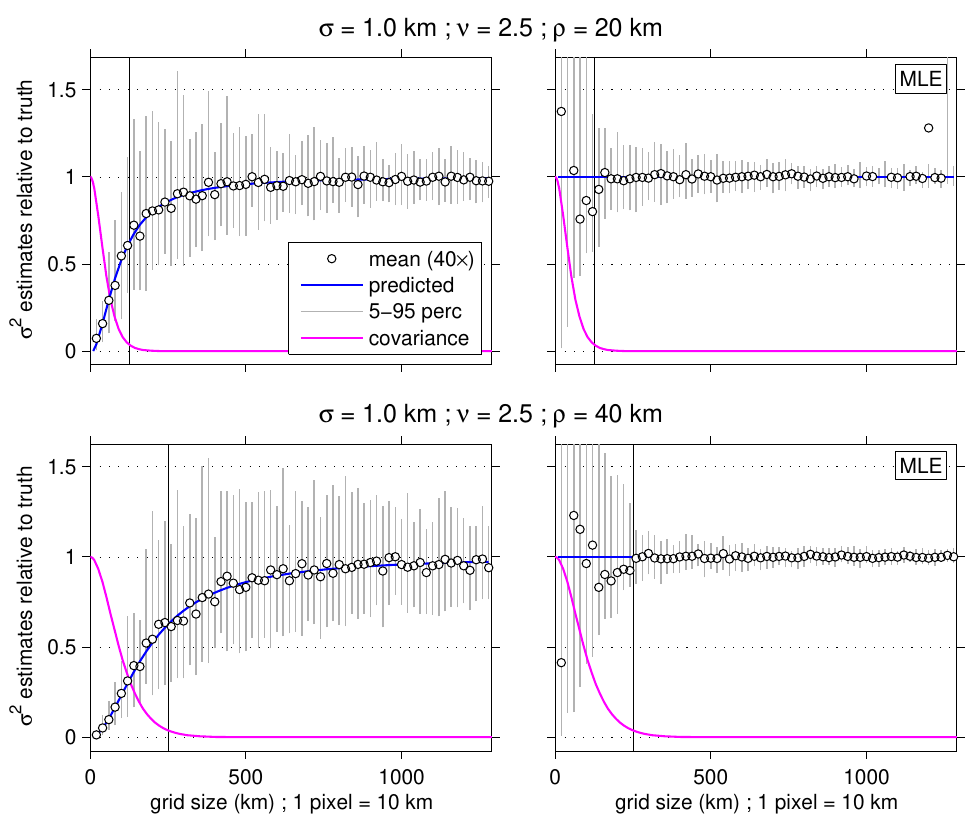}
\caption{\label{poor}The sample variance $\mathrm{s^2}$ is a biased,
  inconsistent, and inefficient estimator for the true process variance
  $\st$. The maximum-likelihood estimator is asymptotically unbiased, consistent
  and efficient. Conducting 40 lattice simulations on differently sized data
  patches, with Mat\'ern parameters $(\sigma^2,\nu,\rho)$ as listed in the
  titles, the left panels show the behavior of the sample variance
  $\mathrm{s^2}$, and the right panels that of the maximum-likelihood variance
  estimator (`MLE'), both normalized by the actual variance $\st$. The gray bars
  span the 5th to 95th percentiles of the estimates at the quoted patch sizes,
  the black open circles are the mean estimates, and the solid blue lines their
  predictions from eq.~(\ref{svar2}), as in Fig.~\ref{varbias}. The magenta
  curves are the scaled spatial correlation functions, with the vertical black
  lines at $2\pi\rho$. The means of the MLE for field sizes smaller than
  $2\pi\rho$ were calculated over the 80th percentile of the estimates.}
\end{figure*}

Fig.~\ref{poor} (left panels) reveals that the estimation variance of $s^2$ is
generally high (relatively speaking), and decaying too slowly (for our taste)
with increasing grid size. In comparison, the quasi-maximum-likelihood estimator
that we develop in the next section has properties that are far more favorable,
as is readily, if proleptically, illustrated by the second suite of experiments
shown in Fig.~\ref{poor} (right panels, marked `MLE'). Saving the details of its
construction for the next section, inspecting the figure reveals that, as soon
as the data patch size exceeds the correlation length of the Mat\'ern process,
the estimation variance of the maximum-likelihood variance estimator is
acceptably low. Moreover, the estimation variance continues to decay at a
pleasing rate, suggestive of its asymptotic unbiasedness.

While the examples thus far may have appeared anecdotal, it is hoped that they
do convincingly hint at the agreeable qualities of the maximum-likelihood
estimators, which we now discuss in more detail.

\section{M~A~X~I~M~U~M~-~L~I~K~E~L~I~H~O~O~D{\hsps}T~H~E~O~R~Y}
\label{secmle}

The material in this section is chiefly inspired by sections 4.3--4.8 and
Appendix~A6 of \cite{Simons+2013}. \cite{Cox+74} remains an excellent reference,
though more modern texts such as \cite{Pawitan2001} and \cite{Severini2001} are
recommended. Our main device is the frequency-domain \cite{Whittle53,Whittle54}
likelihood, modified to acknowledge edge effects by blurring the spectral
density function.

\subsection{Finite large-sample theory}

\cite{Simons+2013} introduced $\mcL\oft$, the likelihood for the Mat\'ern
parameters, which, as is at best acceptable only for large sample sizes, ignores
the blurring in the isotropic spectral density~$\mcS\ofst\ofsk$ as well as the
correlation induced between wavenumbers, in the form
\begin{equation}\label{lik}
\mcL\oft=
-\norml
\left[
\ln \mcS\ofst\ofsk
+\mcS\ofst^{-1}\hsomm\ofsk\hsom|H\ofk|^2\right]
.
\end{equation}

Its first derivatives, with respect to each of the parameters~$\theta\in\btheta$, 
are the components of the unblurred score vector~$\bgamma\oft$, given by
\begin{equation}\label{score}
  \gth\oft
  =\frac{\partial\mcL(\btheta)}{\partial\theta}
  =-\norml\mth\ofsk\left[1- \mcS\ofst^{-1}\hsomm\ofsk\hsom|H\ofk|^2\right],
\end{equation}
with, easily obtained via differentiation of eq.~(\ref{matern2}), and listed as
eq.~(\ref{mst}) in the Appendix, the required expressions 
\begin{equation}\label{mth}
\mth\ofsk=\mcS\ofst^{-1}\hsomm\ofsk\frac{\pl\mcS\ofst\ofsk}{\pl\theta}
.
\end{equation}

Its second derivatives, with respect to two arbitrary
parameters~$\theta,\theta'\in\btheta$, are the components of the Hessian
matrix~$\bF\oft$,
\begin{equation}\label{hessian}
  F_{\theta\theta'}\oft
  =\frac{\partial\gamma_{\theta'}(\btheta)}{\partial\theta}
  =\frac{\partial^2\mcL(\btheta)}{\partial\theta\partial\theta'}
  =-\norml\Bigg[
\frac{\pl \mthp\hsomm\ofsk}{\pl\theta}
+\left\{\mth\ofsk \mthp\hsomm\ofsk
-\frac{\pl \mthp\hsomm\ofsk}{\pl\theta}\right\}
\left\{
\mcS\ofst^{-1}\hsomm\ofsk\hsom|H\ofk|^2
\right\}
\Bigg]
,
\end{equation}
with the nonvanishing derivatives $\plth\mthp(k)$ given as
eqs~(\ref{m1})--(\ref{m6}) in the Appendix.

The negative expectation of $\bF\oft$ is the Fisher matrix~$\bmcF\oft$, which
does not depend on the data as
$\langle\mcS\ofst^{-1}\hsomm\ofsk\hsom|H\ofk|^2\rangle=1$, and thus
\begin{equation}\label{fisher}
  \mcF_{\theta\theta'}\oft
  =-\left\langle\frac{\partial^2\mcL(\btheta)}{\partial\theta\partial\theta'}\right\rangle
  =\norml\mth\ofsk\hsom\mthp\hsomm\ofsk
.
\end{equation}
The inverse of the Fisher matrix is the information matrix,
$\bmcF^{-1}=\bmcJ$. Denoting the true parameter set as~$\btruth$, with elements
$\theta_0$, and the maximum-likelihood estimate as~$\hbt$, with elements $\hth$,
the presumed normality of the Fourier coefficients~(\ref{fourierHw}) implies the
distribution
\begin{equation}\label{summary2}
\sqrt{MN}(\hbt-\btruth) \sim \mcN(\bzero,\bmcF^{-1}\hsomm\oftr)=\mcN(\bzero,\bmcJ\oftr)
,
\end{equation}
from which we will seek to construct $100\times(1-\beta)$ per cent
confidence intervals about the estimates, using the values
$z_{\beta/2}$ at which the standard-normal distribution reaches a
cumulative probability of $(1-\beta/2)$, as follows:
\begin{equation}\label{conflocationtwo}
\hth-z_{\beta/2}\frac{\Jthth\shalf\ofth}{\sqrt{MN}}
\le
\truth
\le
\hth+z_{\beta/2}\frac{\Jthth\shalf\ofth}{\sqrt{MN}}
.
\end{equation}

The relations in this section are theoretical quantities derived by
\cite{Simons+2013} that, strictly speaking, apply only to the `population' case,
on domains of infinite extent, and we note that caveats about consistency apply
for $MN$ large in the fixed-domain asymptotic regime, as we alluded to in
Sec.~\ref{introduction}. While even under increasing-domain asymptotics, there
may be few rigorous results on maximum likelihood estimates for the Mat\'ern
model when the smoothness parameter is treated as unknown, the \textit{(highly)
  significant correlation contribution} introduced by \cite{Guillaumin+2022}
helps illuminate the estimability of, and the identifiability between, the three
parameters, and the separation of different processes under various spatial
sampling mechanisms, which will enable pursuing these issues further.  In the
`sample' case of discretized, windowed likelihood analysis, we follow
\cite{Simons+2013} and \cite{Guillaumin+2022} in replacing the Mat\'ern spectral
density $\mcS\ofst\ofsk$ in eq.~(\ref{lik}) with a suitably blurred version,
$\Sbar\ofst\ofsk$, to acknowledge the effects of finite sampling. In what
follows, we will explore the implications of sampling and bounding on the
uncertainty estimates of the parameters.

When using the Whittle likelihood as defined in eq.~(\ref{lik}) to produce
estimators for time series, the behavior of the spectral estimates is very well
understood, see for example \cite{Dzhaparidze+83}. Exactly how the Whittle
likelihood is formulated may vary---in terms of an integral or a sum, with the
log covariance term included or not \cite[]{Hosoya+1982}, and with and without
tapering \cite[]{Dahlhaus1984}. In two dimensions and higher, however, it is
recognized that the boundary effects become dominant, which has led to
alternative formulations \cite[]{Guyon82,Deb+2017,Guillaumin+2022}.
\cite{Grainger+2025} treat the non-parametric asymptotics of (even
non-separable) tapers coherently. Here we follow \cite{Guillaumin+2022} in
producing estimators from the forthcoming spectral eq.~(\ref{blik}), which
removes bias from boundary effects, can be tapered, maintains the statistical
efficiency of a spatial likelihood, and remains computationally
competitive. Should the parameters of the model become sample-size dependent, no
existing theory is available yet.

\subsection{Finite sampled data: heuristics}

\begin{figure*}\centering
\includegraphics[width=0.75\textwidth,angle=-0]{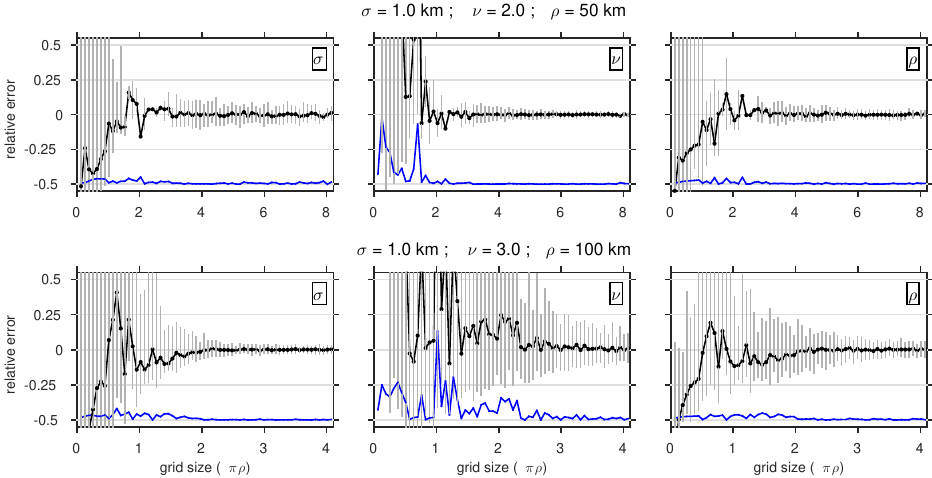}
\caption{\label{mle}Behavior of the relative error of the Whittle
  maximum-likelihood estimators of the Mat\'ern parameters $(\sigma^2,\nu,\rho)$
  for the two sets of true values listed in the titles, conducted on square
  lattices growing in size from $M=N=2$ to $M=N=128$ pixels of size
  $\Dx=\Dy=10~\mathrm{km}$, quoting $M\Delta x$ in multiples of $\pi\rho$. Gray
  bars cover the 5th through 95th percentiles of the estimates for each set of
  up to 40 simulations.  Black filled circles are the means of the estimates,
  computed over the 80th percentile of the sets for fields whose linear
  dimension $M\Dx<2\pi\rho$, but over the full set of up to 40~estimates beyond
  that size. With growing grid size, the estimates reveal themselves to be less
  biased with shrinking variance. The blue line tracks the absolute value of the
  relative error multiplied by $\sqrt{MN}$, offset by $-0.5$ for visual
  clarity.}
\end{figure*}

For sampled data, \cite{Simons+2013} defined the likelihood of observing the
data~$\mcH\ofx$ under the spectral model~(\ref{matern2})--(\ref{thetaS})
parameterized by $\btheta$. This `blurred likelihood' $\bar{\mcL}\oft$ instead
of $\mcL\oft$, is given in terms of the Fourier coefficients of the gridded and
windowed data, $H\ofk$ in eq.~(\ref{fourierHw}), and of the blurred isotropic
spectral density, $\Sbar\ofst\ofk$ of eq.~(\ref{varHk3}), summed over all
wavenumbers in the grid~(\ref{kgrid}), by %the relation
\begin{equation}\label{blik}
\bar{\mcL}\oft=
-\norml
\left[
\ln \Sbar\ofst\ofk
+\Sbar\ofst^{-1}\hsomm\ofk\hsom|H\ofk|^2\right]
.
\end{equation}
Compare eq.~(\ref{blik}) with eq.~(\ref{lik}): the only difference is the
acknowledgment of the spectral blurring effect of sampled data.
Eq.~(\ref{blik}) is the quantity that we maximize, under positivity constraints,
for the parameter vector~$\btheta$, thereby defining the
quasi-maximum-likelihood estimate~$\hbt$ to be obtained by solving for the
`score' vector~$\bbg\oft$ of numerical first derivatives of the blurred
likelihood, in the sense
\begin{equation}\label{sats}
\bbg\ofth=\bzero
.
\end{equation}
Satisfying eq.~(\ref{sats}) to find $\hbt$, for example by an iterative function
minimization routine, requires repeated evaluation of the spectral
density~(\ref{matern2}) on the grid~(\ref{kgrid}), with the blurring implemented
convolutionally (on a refined and subsequently reinterpolated grid) or else
exactly, as in eq.~(\ref{varHk3}). Without entering into more details for now,
Fig.~\ref{mle} shows the results of a suite of experiments conducted to
illustrate the performance of our numerical method that recovers each of the
three Mat\'ern parameters $(\sigma^2,\nu,\rho)$, as a function of grid size
(measured in terms of the correlation length $\pi\rho$). As the right-hand side
panels of Fig.~\ref{poor} showed for the variance estimate, the
maximum-likelihood estimates are very well-behaved, from about the point where
the grid size reaches a linear dimension of about $\pi\rho$. The procedure
almost surely yields low-variance and practically unbiased estimates from a grid
size of about $2\pi\rho$ onward, as Fig.~\ref{mle} shows empirically.

The mathematical form and geometry of the blurred likelihood function are what,
fundamentally, controls the observed behaviors. \cite{Simons+2013} only
considered convolutional approaches to blurring, and were limited in their
ability to acknowledge the spectral interaction induced by the applied data
windows on the parameter covariance estimates. The first and second derivatives
of the blurred likelihood are \textit{not} simply obtained by substituting
$\Sbar\ofst\ofsk$ for $\mcS\ofst\ofsk$ in eqs~(\ref{score}) and~(\ref{hessian}),
since the factors~$\mth\ofsk$ that appear in the expressions for the score and
the Hessian are \emph{analytical} derivatives that involve the
\textit{unblurred} spectral density~$\mcS\ofst\ofsk$. Replacing $\mcS\ofst\ofsk$
by $\Sbar\ofst\ofsk$ in eq.~(\ref{score}) yields \emph{reasonable}
approximations for the likelihood derivatives, which compare \emph{relatively}
favorably to their numerical counterparts---especially for large sample
sizes. Most numerical optimization routines will be able to maximize
eq.~(\ref{blik}), solving eq.~(\ref{sats}) without being given explicit
expressions for the score and the Hessian. However, in order to derive accurate
confidence intervals on our estimated parameters, we do need access to the
expected value of the second-order derivatives of the \textit{actual} likelihood
that is being maximized.

Numerical experiments and theoretical considerations along the lines suggested
in their Appendix~A8 tempted \cite{Simons+2013} into concluding that
eq.~(\ref{conflocationtwo}) could be used to construct confidence intervals for
the solutions of eq.~(\ref{sats}) in our present case of univariate
two-dimensional analysis. Under the viewpoint espoused in their
eqs~(A84)--(A87), the blurred spectrum is an \textit{additive} correction term
(small for smoothly varying spectra) away from the original. In this framework,
neglecting to blur the Fisher matrix (not to mention neglecting wavenumber
correlations) was believed to have an altogether negligible effect on the
estimation variances based on its inverse, even if blurring the likelihood is
absolutely essential to arrive at the estimate in the first place. However, the
ability of the \textit{unblurred} Fisher matrix to help predict the variance of
the parameters derived via maximization of the \textit{blurred} likelihoods
turns out to be poor, especially as concerns the variance and correlation
parameters $\st$ and $\rho$. The unblurred expression, eq.~(\ref{fisher}), of
the Fisher matrix provides an asymptotic but ultimately inadequate match to the
average of the numerical Hessian for real-world sampling scenarios.

Not accounting for wavenumber correlation proved to be another stumbling block.
\cite{Simons+2013} conceived of approximations to account for wavenumber
correlation involving a \textit{multiplicative} correction term (their
eqs~A56--A58). For very large sample sizes this correction term approaches
unity. Contrary to the optimism they expressed, uncertainty estimates for the
maximizers of eq.~(\ref{blik}) that rely on eq.~(\ref{conflocationtwo}) are
inadequate for all but the largest sample sizes. A heuristic way of determining
the estimation variance for the recovered parameters when actual data are being
investigated is to generate synthetics with features identical to those of the
gridded data, from models with Mat\'ern parameters given by previously obtained
solutions, then estimating their parameters a number of times, and learning from
their distribution what the likely uncertainty ranges for the parameters of the
actual data patches might be, as in Fig.~\ref{mle}. However justified, little
transferable knowledge is gained in the process, and the procedure is cumbersome
and time-consuming.

\cite{Guillaumin+2022} showed the way forward by further developing the theory
of likelihood analysis for finite sampled data, on which we rely to develop the
practical methods offered in the next sections. They include the ability to
calculate uncertainty estimates on the parameters from first principles. The
next section provides a complete description of the entire workflow.

\subsection{Finite sampled data: full theory}\label{fourthree}

For sampled data, the likelihood involves~$\Sbar\ofst\ofk$, the blurred spectral
density (eqs~\ref{specblur2} and~\ref{varHk3}), and the modified periodogram of
the data~$\mcH\ofx$,
\begin{equation}\label{blik2}
\bar{\mcL}\oft=
-\norml
\left[
\ln \Sbar\ofst\ofk
+\Sbar\ofst^{-1}\hsomm\ofk\hsom|H\ofk|^2\right]
,
\end{equation}
where $H\ofk$ is the windowed Fourier transform of the data~$\mcH\ofx$, for an
arbitrary unit-normalized window~$w\ofx$, repeating eq.~(\ref{fourierHw}),
\begin{equation}\label{fourierH2} 
H\ofk
\equiv
\frac{1}{2\pi}\nff\sum_\xb w\ofx\mcH\ofx e^{-i\kb\cdot\xb}
,
\end{equation}
and the Mat\'ern spectral density~$\mcS\ofst$ whose parameters we aim to recover
(see eq.~\ref{matern2}) is exactly blurred (hence no longer isotropic, see
Fig.~\ref{square}) to account for finite-sample effects via the intermediary of
the isotropic spatial Mat\'ern covariance~$\mcC\ofst\ofsy$, see eq.~(\ref{Kr}),
repeating eq.~(\ref{varHk3}),
\begin{equation}\label{interm}
\Sbar\ofst\ofk=\frac{1}{(2\pi)^2}\nffs
\sum_\yb W(\yb) \hsom\mcC\ofst\ofsy e^{-i \kb \cdot \yb}
=\var\left\{H\ofk\right\}
,
\end{equation}
and with~$W$ the autocorrelation of the sampling window, obtainable via FFT,
that is, repeating eq.~(\ref{wyb}) over the relevant grid (lags),
\begin{equation}
W(\yb)=\sum_\xb w\ofx w(\xb+\yb)
.
\end{equation}
The solution~$\hbt$ is found by maximization of eq.~(\ref{blik2}), requiring the
vanishing of the score $\bnabla_{\!\ofst}{\bar{\mcL}\oft}=\bbg\oft$, whose
components are given by
\begin{equation}\label{bgth}
\bgth\oft=-\norml\bmth\ofk\left[1-\Sbar\ofst^{-1}\hsomm\ofk\hsom|H\ofk|^2\right],
\end{equation}
with the blurred equivalents to eq.~(\ref{mth}) again obtained exactly via the
intermediary of eq.~(\ref{interm}) as
\begin{equation}\label{mthbar}
  \bmth\ofk=
  \Sbar\ofst^{-1}\hsomm\ofk\frac{\pl\Sbar\ofst\ofk}{\pl\theta}=
  \frac{\Sbar\ofst^{-1}\hsomm\ofk}{(2\pi)^2}\nffs
  \sum_\yb W(\yb)\hsom\frac{\pl\mcC\ofst\ofsy}{\pl\theta} e^{-i \kb \cdot \yb}
.
\end{equation}
The requisite derivatives of the spatial covariance $\pl_\theta\mcC\ofst$ are
obtained via differentiation of eq.~(\ref{Kr}) and listed as
eqs~(\ref{Cst})--(\ref{Crho}) in the Appendix.

The components of the Fisher matrix $\bbF\oft$, with respect to two arbitrary
Mat\'ern parameters $\theta,\theta'\in\btheta$, are now given by
\begin{equation}\label{fisherb}
\bar\mcF_{\theta\theta'}\oft=\norml\bmth\ofk\hsom\bmthp\hsomm\ofk
.
\end{equation}

As \cite{Simons+2013} (their eq.~138), but now following
\cite{Guillaumin+2022} (their eq.~36), the parameter estimation variance,
\begin{equation}\label{magic}
  \cov\ofth\approx \bar{\bmcF}^{-1}\oftr\,\cov\{\bbg\oftr\}\,\bar{\bmcF}^{-1}\oftr
  ,
\end{equation}
where in practice we substitute $\boldsymbol{\hat{\theta}}$ for
$\boldsymbol{\theta_0}$, to which the estimator converges in probability, 
requires the additional calculation of the covariance of the score
\textit{without} neglecting the correlation between wavenumbers,
\begin{equation}\label{covgammaS}
\cov\big\{\bgth,\bgthp\big\}=\normltu \bmth\ofk
\,\frac{\cov\{|H\ofk|^2,|H\ofkp|^2\}}{\Sbar\ofst\ofk\Sbar\ofst\ofkp}
\hsom\bmth\hsomm\ofkp
,
\end{equation}
see \cite{Guillaumin+2022} (their eq.~37), and compare \cite{Simons+2013} (their
eq.~A54, which should have quoted blurred quantities). Eq.~(\ref{covgammaS})
implies that we require the covariance of the windowed periodogram $|H\ofk|^2$,
which under standard theory using the \cite{Isserlis18} theorem for Gaussian
processes is \cite[][their eq.~A57]{Percival+93,Stein95,Simons+2013}
\begin{equation}\label{toomet}
\cov\big\{|H\ofk|^2,|H\ofkp|^2\big\}=
\left|\cov\{H\ofk,H^*\hsomm\ofkp\}\right|^2+
\left|\cov\{H\ofk,H\ofkp\}\right|^2
.
\end{equation}

\subsection{Covariance of the estimates}

To obtain $\cov\ofth$ in eq.~(\ref{magic}), $\cov\{\bbg\ofst\}$ in
eq.~(\ref{covgammaS}) can be stochastically approximated by repeated simulation
of the random field $\mcH\ofx$, see Sec.~\ref{correlation}, and calculating
$\bbg\oft$ of eq.~(\ref{bgth}) via numerical differentiation. Used inside
eq.~(\ref{magic}), this readily produces acceptable approximations for the
parameter covariance in all cases we considered.  Eq.~(\ref{covgammaS}) can also
be calculated \textit{exactly} using one of two methods.

\begin{figure*}\centering
\includegraphics[width=\textwidth,angle=0,trim=0cm 0cm 0cm 0cm,clip]{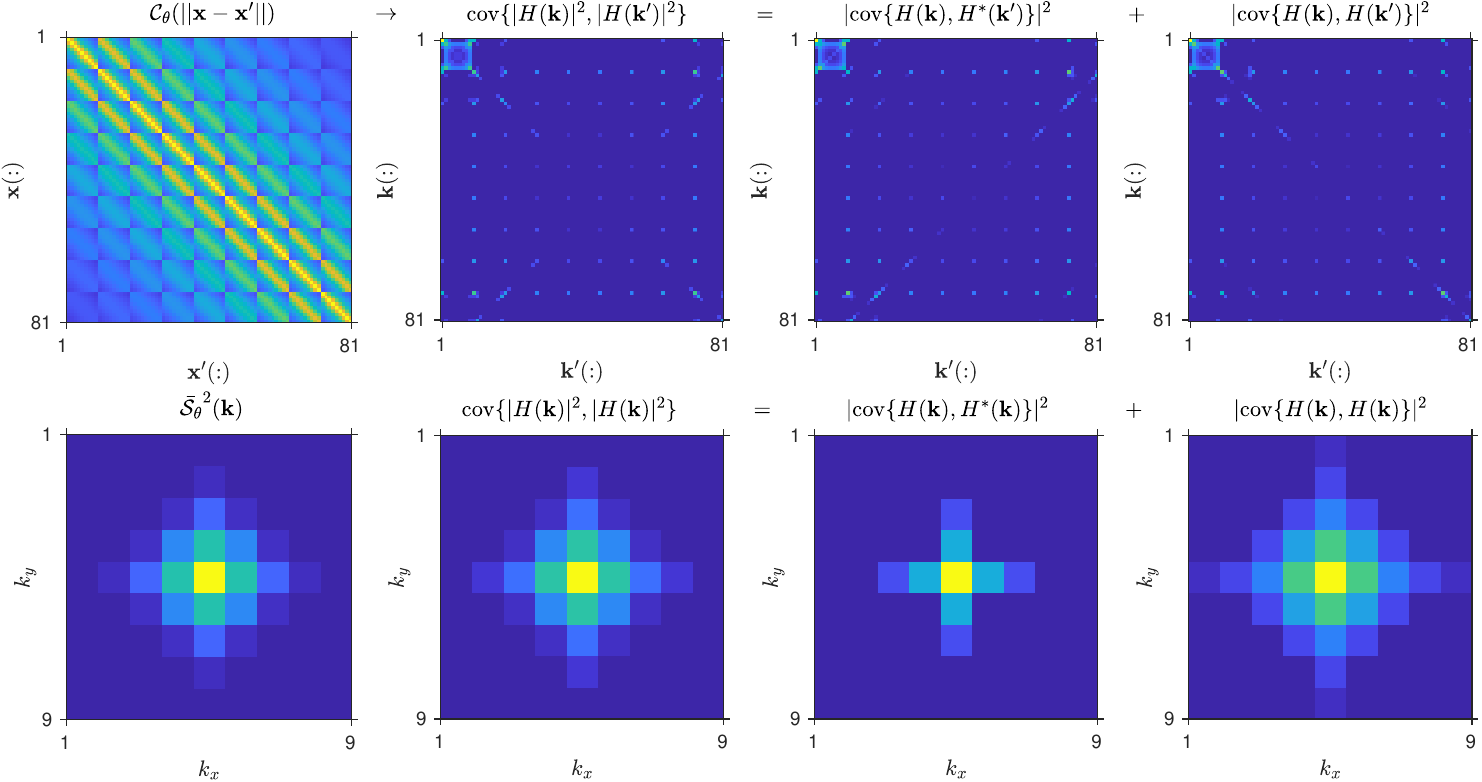}
\caption{\label{bluabl}Sampling and windowing effects on the covariance between
  periodograms: blurring and correlation illustrated for a Mat\'ern model with
  parameters $\btruth=(\st,\nu,\rho)=(1,0.5,2)$ on a grid of $M=N=9$ pixels of
  size $\Dx=\Dy=1$ in arbitrary units. (\textit{Top row}) The spatial covariance
  $\mcC\ofst(\xb,\xbp)$ on the grid that gives rise to the periodogram
  covariance $\cov\big\{|H\ofk|^2,|H\ofkp|^2\big\}$, which is the sum of two
  (pseudo-)covariance terms, see eq.~(\ref{toomet}). The off-diagonal terms
  express the correlation between periodograms at different frequencies; the
  diagonal terms embody their blurring. (\textit{Bottom row}) The squared
  blurred spectral density~$\bar{\mcS}\ofst^2\hsomm\ofk$ is close to the
  variance of the periodograms $\var\big\{|H\ofk|^2\}$, as shown in the last
  three panels, which are identical to the diagonals of the panels above them,
  wrapped into a two-dimensional spectral grid.}
\end{figure*}

The first is by direct calculation of the covariance between the
windowed periodograms in eq.~(\ref{toomet}), rewritten \cite[]{Walden+94} as
\begin{align}\label{toometu}
\cov\big\{|H\ofk|^2,|H\ofkp|^2\big\}&=
\left|\langle
H\ofk H\ofkp
\rangle
\right|^2+
\left|\langle
H\ofk H^*\hsomm\ofkp
\rangle
\right|^2\\
&=\label{toometv}
\big| \sQ\hsom \langle \mcH\ofx\mcH\ofxp \rangle \hsom\sQ' \big|^2+
\big| \sQ\hsom \langle \mcH\ofx\mcH\ofxp \rangle \hsom{\sQ^*}' \big|^2\\
&=\label{toomets}
\big| \sQ\hsom \mcC\ofst \hsom\sQ' \big|^2+
\big| \sQ\hsom \mcC\ofst \hsom{\sQ^*}' \big|^2
,
\end{align}
where, abusing notation ever so slightly, $\sQ$ is the discrete-Fourier
transform matrix operator that transforms windowed spatial-domain observations
to the spectral domain, $\sQ'$ and ${\sQ^*}'$ its transpose and conjugate
transposes, respectively, and $\mcC\ofst$ the Mat\'ern autocovariance sequence
at all available lags. Fig.~\ref{bluabl} illustrates this computational
procedure for a spatial observation grid with a unit-window taper.  The top row
shows the covariance sequence $\mcC\ofst(\|\bx-\bxp\|)$ with all two-dimensional
lag-distances unwrapped along the two axes labeled $\bx(:)$ and $\bxp(:)$,
followed by the covariance of the periodograms as the sum of the individual
(pseudo-)covariance \cite[]{Neeser+93} terms as they appear in
eq.~(\ref{toomet}). The bottom row shows the $\kb=\kbp$ diagonal elements of
each of the latter, wrapped as $k_x$ and $k_y$ about the zero wavenumber in the
center, following the independently (via eq.~\ref{interm}) calculated square of
the blurred spectral density, as labeled. The main diagonal of the periodogram
covariance, its variance
$\cov\big\{|H\ofk|^2,|H\ofk|^2\big\}=\var\big\{|H\ofk|^2\}$, matches the square
of the blurred spectral density~$\bar{\mcS}\ofst^2\hsomm\ofk$.

The second implementation for calculating the covariance of the blurred and
correlated score is by considering the terms in eq.~(\ref{toometu}). Returning
to the integral form of eq.~(\ref{specblur}), rewritten with the explicit
acknowledgment of a window function $w(\xb)$, which may not be unitary, used as
a subscript, we write, integrating over the Nyquist plane of frequencies, for
the case of the second term of eq.~(\ref{toometu}),
\begin{equation}\label{specblur3}
  \langle H\ofk H^*\hsomm\ofkp\rangle=
  \intnyqk D_K(\kb-\kb'')D_K^*(\kb'-\kb'')\hsom\mcS\ofst(\kb'')\dbk''
  \where
  D_K\ofk=\frac{1}{2\pi}\nff\sum_\xb w\ofx e^{-i\kb\cdot\xb}
  .
\end{equation}
At a fixed distance between wave vectors~$\Delta\kb=\kb-\kbp$,
eq.~(\ref{specblur3}) amounts to a simple convolution of the theoretical
spectral density with a kernel that is the product of two relatively offset
Fourier series, $D_K(\kb)D_K^*(\kb+\Delta\kb)$, which we can in turn calculate
as a Fourier series via the convolution theorem. When $\Delta\kb=\bzero$,
without offset, we recover the variance of the windowed Fourier coefficients as
the expectation of the periodogram via convolution of the spectral density with
the modified Fej\'er kernel $\left|D_K(\kb)\right|^2$. Analogously with
eq.~(\ref{interm}) we compute
\begin{equation}\label{specblur4}
  \langle H\ofk H^*\hsomm\ofkp\rangle=
  \frac{1}{(2\pi)^2}\nffs \sum_\yb W(\yb,\Delta\kb)\,\hsom\mcC\ofst\ofsy\hsom e^{-i \kb \cdot \yb}
  ,\with \Delta\kb=\kb-\kbp,
\end{equation}
adapting the autocorrelation of the energy-normalized analysis window as follows:
\begin{equation}
W(\yb,\Delta\kb)=\sum_\xb w\ofx w(\xb+\yb)\hsom e^{i\Delta\kb\cdot(\xb+\yb)}
.
\end{equation}
Proceeding similarly, after reindexing and taking into account relevant
symmetries for the pseudocovariance first term in eq.~(\ref{toometu}), yields
\begin{equation}\label{specblur5}
  \langle H\ofk H\ofkp\rangle=
  \intnyqk D_K(\kb-\kb'')D_K(\kb'+\kb'')\hsom\mcS\ofst(\kb'')\dbk''
  .
\end{equation}
This procedure requires one Fast Fourier Transform per diagonal increasingly
offset from the main diagonal in $(\kb,\kbp)$ space, further multiplied with
$\bmth\ofk/\Sbar\ofst\ofk$ and $\bmth\ofkp/\Sbar\ofst\ofkp$ and normalized into
eq.~(\ref{covgammaS}). Like the procedure in
eqs~(\ref{toometu})--(\ref{toomets}), it is exact \cite[as opposed to Riemann
  discretization of eqs~(\ref{specblur3}) and~(\ref{specblur5}), as originally
  proposed by][]{Guillaumin+2022}.

Assembling all the various pieces, we now have three different methods to obtain
the desired estimation variance in eq.~(\ref{covgammaS}): likelihood gradient
sampling, direct calculation via the DFT matrix, and per-diagonal Fourier
transformation of the spatial covariance. The first is fast and has shown good
behavior in all of our tests, see Sec.~\ref{numerics}. The second and third
produce identical results, with the matrix implementation being memory
intensive, and the per-diagonal sequential calculation still computationally
heavy as it calls on $MN$ FFTs, each of which requiring $MN\log(MN)$
operations. As shown in Fig.~\ref{bluabl}, not all interaction terms are large,
which has inspired randomized approaches to capture their contributions
\cite[]{Stein+2004}, e.g., via Markov-Chain Monte Carlo
\cite[]{Metropolis+1949,Hastings1970} or importance sampling. Managing the
interplay of the stochastic process being sampled and the characteristics of the
sampling strategy is amenable to formal treatment via the \textit{significant
  correlation contribution} as developed by
\cite{Guillaumin+2017,Guillaumin+2022}.

\subsection{Synthetic numerical examples}\label{numerics}

While all of our results discussed thus far hold even when the analysis window
$w\ofx$ is an irregularly bounded and possibly incompletely sampled subset of
the regular grid~(\ref{xgrid}), in the remainder of this paper we focus on
completely sampled $M\times N$ rectangular grids.

\begin{figure*}\centering
\includegraphics[width=0.65\textwidth,angle=0,trim=0cm 1.25cm 0cm 2.1cm,clip]{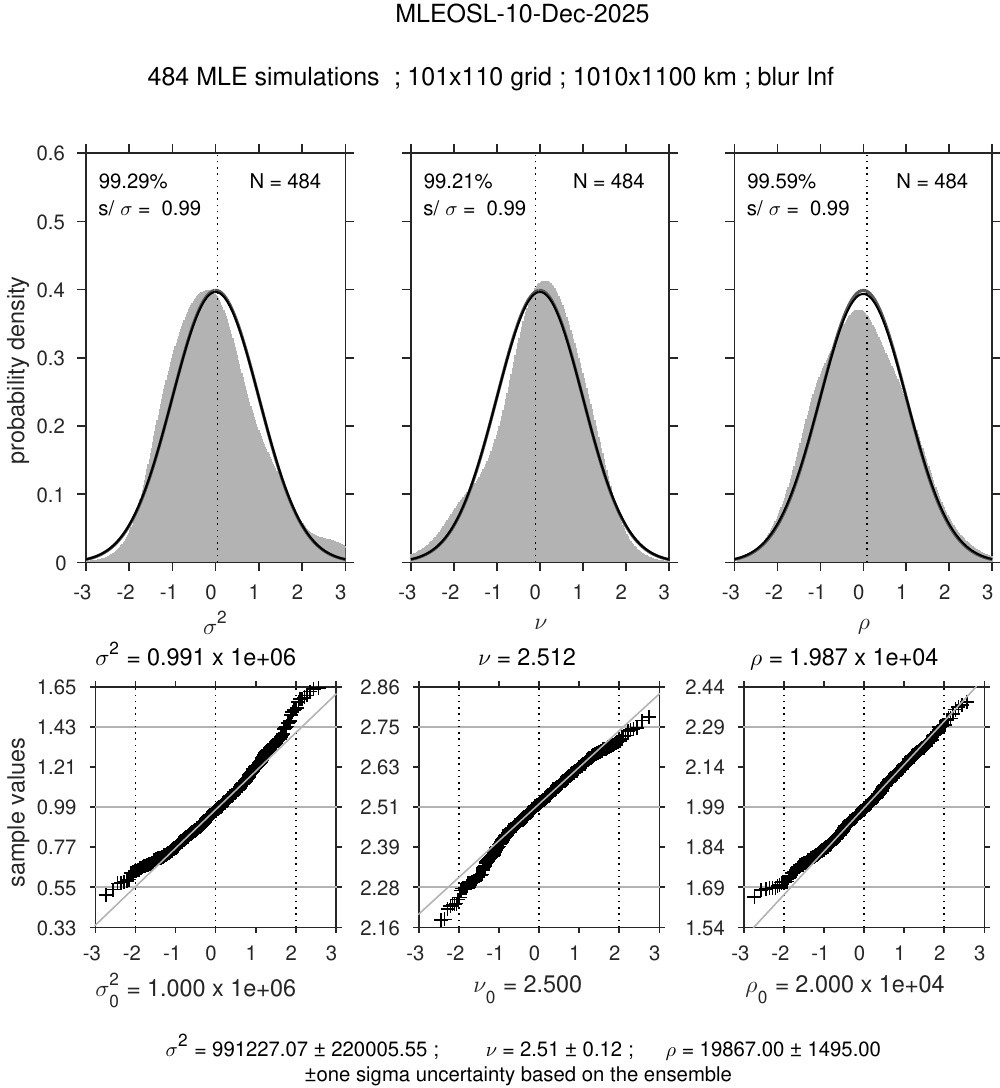} 
\caption{\label{mleosl}Mat\'ern parameter maximum-likelihood estimation
  statistics for 484 spatial-covariance embedding simulations carried out on a
  101$\times$111 grid, with spacings $\Dx=\Dy=10$~km, and true values of
  $\sigma^2_0=1$~km$^2$, $\nu_0=2.5$, $\rho_0=20$~km, recovered via maximization
  of the exactly blurred uncorrelated likelihood~(\ref{blik2}). The estimates
  average to $\st=0.91\pm0.22$~km$^2$, $\nu=2.51\pm0.12$,
  $\rho=19.87\pm1.509$~km, quoting one observed standard deviation. In the top
  row, the thick dark gray line overlying the light gray shaded smooth
  kernel-density estimate of the ensemble of simulation and estimation outcomes
  is the Gaussian derived from the covariance of the ensemble. The (nearly
  entirely overlapping) thick black line centered on the mean estimate is based
  on the covariance exactly calculated from eq.~(\ref{magic}) evaluated at the
  truth, indicated by the dotted vertical line in the top panels. In the bottom
  row, quantile-quantile plots compare the empirical distribution to a normal
  distribution standardized using the mean and standard deviation of the
  ensemble.}
\end{figure*}

Fig.~\ref{mleosl} is a representative illustration of the behavior of the
estimator~$\hbt$ of eqs~(\ref{blik})--(\ref{blik2}), in each of the Mat\'ern
parameters $\st$, $\nu$, and~$\rho$. We conducted five hundred inversions (484
of which successfully yielded estimates) on independently generated simulations
(via circulant embedding of the spatial covariance), and studied the empirical
distribution of the estimates, compared to the theoretical expression of
eq.~(\ref{magic}) using the exact method of
eqs~(\ref{toometu})--(\ref{toomets}). Invariably, the estimates are nearly
unbiased, and nearly universally Gaussian distributed, as can be seen from the
histograms and the quantile-quantile plots.  The top row shows the smoothly
estimated \cite[]{Botev+2010} standardized probability density functions
(shaded) of the values recovered in this experiment of sample size~$101\times
111$. The abscissas were truncated to within $\pm$3 of the empirical standard
deviation about the origin at the mean estimate; the percentage of the values
captured by this truncation is listed in the top left of each graph. The thick
gray line summarizes the empirical distributions as a Gaussian with the observed
mean and variance in this set of simulations. The thick black line uses the
corresponding theoretical values calculated from eq.~(\ref{magic}). The ratio of
the empirical to the theoretical standard deviation is shown listed as
$\mathsf{s}/\sigma$ for each of the parameters.  The dotted black line marks the
true parameter values. The bottom row shows the quantile-quantile plots of the
theoretical (abscissa, horizontal) versus the empirical (ordinate, vertical)
distributions. The averages of the recovered values $\st, \nu$, and $\rho$ are
listed at the top of the second row of graphs. The true parameter values
$\sigma^2_0, \nu_0$, and $\rho_0$ are listed at the bottom. We attribute the
slight asymmetry of the histograms and the small bends at the edges of the
quantile-quantile plot to small-sample effects on minimum/maximum order
statistics, and to all Mat\'ern parameters being necessarily positive.

\begin{figure*}\centering
\rotatebox{0}{\includegraphics[width=0.79\textwidth,angle=0,trim=0cm 1.75cm 0cm 2.25cm,clip]{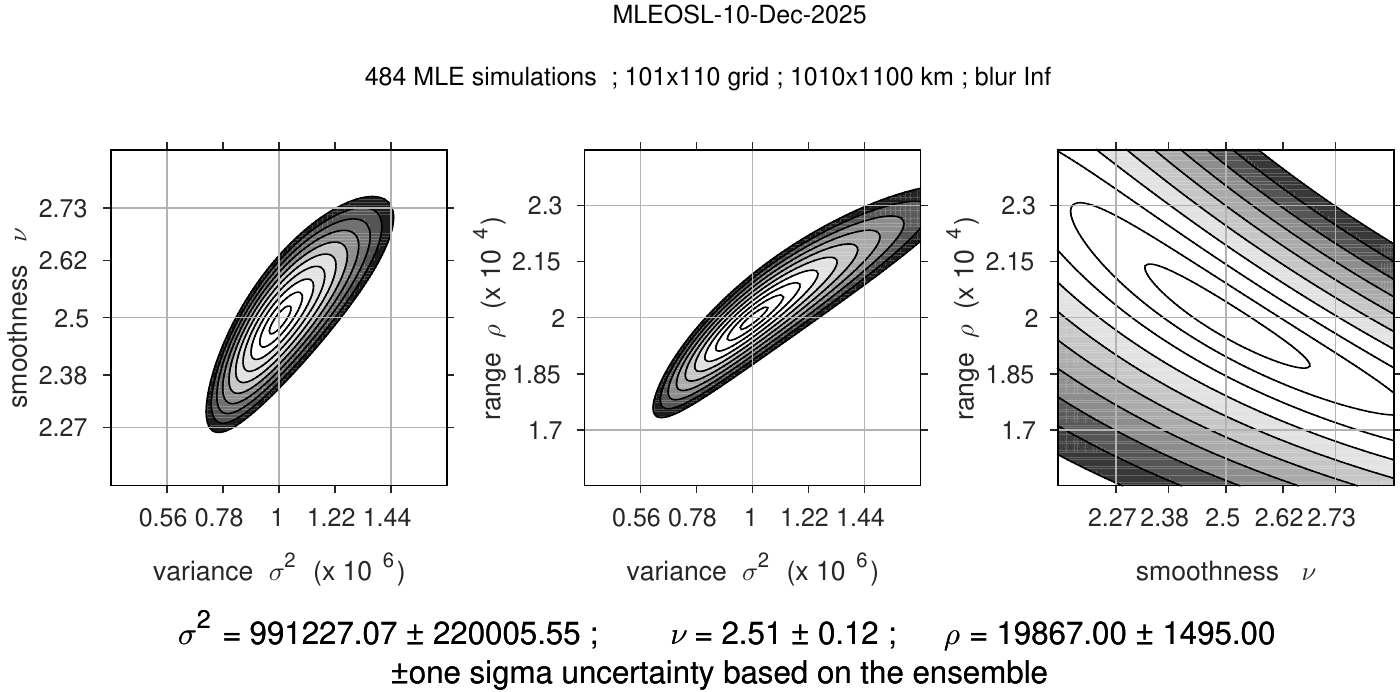}}
\caption{\label{ee}An example of a likelihood~(eq.~\ref{blik2}) surface for
  simulated data from the experiment reported in Fig.~\ref{mleosl}, shown for
  one estimate very close to the truth. Every two-dimensional panel is a slice
  at the value of the third estimated parameter. We draw and shade ten contours
  in equal increments, from the likelihood at the estimate in the center, to the
  likelihood of the parameter pair two observed standard deviations out, in the
  positive quadrant of the ensemble.}
\end{figure*}

Fig.~\ref{ee} is a rendering of the likelihood~(\ref{blik2}) that is being
navigated towards an estimate. We purposely picked a data set from the series of
five hundred runs that yielded an estimate very close to the truth in order to
clean up the axis labeling, but the likelihood surfaces are typically very
similar. Every panel occupies $\pm$3 of the empirical standard deviations around
the estimate. We found ten parameter value pairs evenly spaced out to 2$\times$
the empirical standard deviations from the estimate, and calculated their
likelihoods to use as the shaded contours.

%%%%%%%%%%%%%%%%%%%%%%%%%%%%%%%%%%%%%%%%%%%%%%%%%%%%%%%%%%%%%%%%%%%%%%%%%%%%%%%%
\begin{figure*}\centering
\rotatebox{0}{\includegraphics[width=0.79\textwidth,angle=0,trim=0cm 1.75cm 0cm 2.25cm,clip]{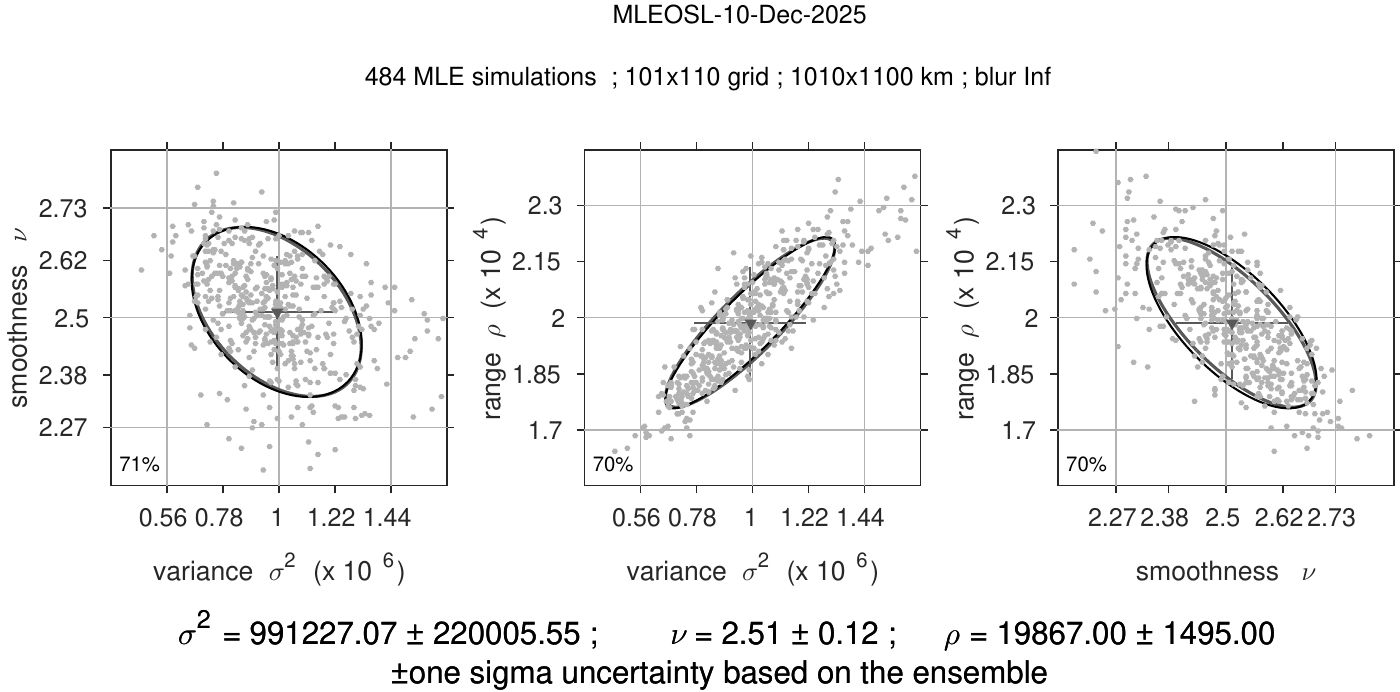}}
\caption{\label{bb}Maximum-likelihood estimation statistics for the ensemble of
  484 simulation and recovery experiments reported in Fig.~\ref{mleosl}. The
  mean estimate is highlighted by the gray triangle and two observed standard
  deviations marked gray by lines. The heavy gray ellipse is the 68\% confidence
  region based on the ensemble. The heavy black ellipse is the predicted 68\%
  confidence region based on the covariance predicted from eq.~(\ref{magic}),
  shown and evaluated at the mean estimate, which is a close match for the set.}
\end{figure*}
%%%%%%%%%%%%%%%%%%%%%%%%%%%%%%%%%%%%%%%%%%%%%%%%%%%%%%%%%%%%%%%%%%%%%%%%%%%%%%%%
\begin{figure*}\centering
\rotatebox{0}{\includegraphics[width=0.55\textwidth,angle=0,trim=0.0cm 0.75cm 0cm 0.5cm,clip]{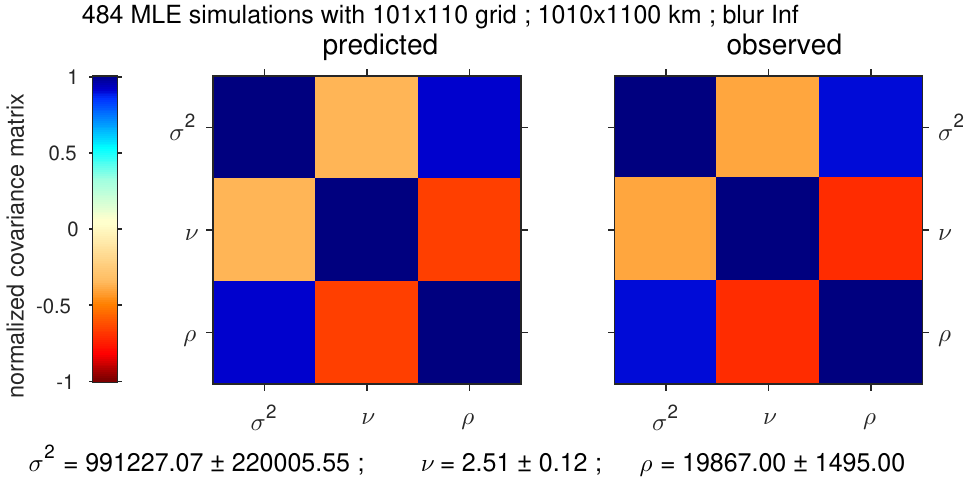}}
\caption{\label{cc}Comparison of the covariance predicted via eq.~(\ref{magic})
  and the covariance observed on the set of experiments reported in
  Figs~\ref{mleosl} and~\ref{bb}. Shown are the relevant correlation matrices
  between the estimators for the three Mat\'ern parameters $\st$, $\nu$, and
  $\rho$, highlighting the relatively strong trade-off between $\nu$ and
  $\rho$.}
\end{figure*}

Fig.~\ref{bb} summarizes the same experiment as cross plots of the recovered
parameters with their empirical summaries \cite[]{Aster+2018} as the ellipsoidal
68\% confidence region (thick gray line), and showing the equivalent error
ellipses from the theoretically predicted covariances (thick black line), which
match the observations closely, albeit not perfectly. Every panel shows all
pairs of parameter estimates (gray circles), with the mean marked as a gray
triangle, and a cross indicating twice the standard deviation for each of the
estimates. In the lower left of each graph we display the percentage of the
estimates that falls within the error ellipse calculated for the observed
covariance. %Nominally, these would all be 68\%.

Fig.~\ref{cc} is an alternative graphical rendition of the theoretically
predicted and empirically observed parameter-estimate correlation matrix, that
is, their covariance matrix normalized to unity along the diagonal. Parameter
trade-offs are inherent in the Mat\'ern description (see, for example,
eq.~\ref{matern0}). The theoretically calculated correlations are
$\{\st,\nu\}=-0.3714$, $\{\st,\rho\}=0.8906$, and $\{\nu,\rho\}=-0.6701$, which
are good predictors for those observed, as the empirical values are
$\{\st,\nu\}=-0.4195$, $\{\st,\rho\}=0.8772$, and $\{\nu,\rho\}=-0.7259$,
respectively.

\subsection{Analysis of residuals}\label{residuals}

The terms $\mcS\ofst^{-1}\hsomm\ofsk\hsom|H\ofk|^2$ and
$\Sbar\ofst^{-1}\hsomm\ofk\hsom|H\ofk|^2$ that have appeared above in the
expressions for the likelihoods~(\ref{lik}), (\ref{blik2}), and their
derivatives~(\ref{score}) and~(\ref{bgth}) contain the ratio of the observed
periodogram of the data to the (blurred) spectral density predicted under the
model. Since the spectral density is a `scale' parameter (as opposed to a
`location') this ratio has the usual interpretation as a measure of misfit. We
have previously noted that if the Gaussian model fits, the expectation of this
quadratic is $\langle\Sbar\ofst^{-1}\hsomm\ofk\hsom|H\ofk|^2\rangle=1$. Here we
follow \cite{Simons+2013} to maintain that twice this quantity, a quadratic form
of Gaussian random variables, given that $\mcH\ofx$ and thus $H\ofk$ are assumed
to be Gaussian and uncorrelated, should be a chi-squared random variable with
two degrees of freedom,
\begin{equation}\label{Xofk}
X\ofst\ofk=\Sbar\ofst^{-1}\hsomm\ofk\hsom|H\ofk|^2\sim\chi^2_2/2 .
\end{equation}
Equipped with this knowledge we can examine how closely the ratios $X\ofst\ofk$,
i.e., the `residuals', follow the distribution~(\ref{Xofk}), and use the match
or lack thereof as a basis to accept or reject the model that the data are
indeed given by a Mat\'ern process of the specified parameters.

It is imprudent to ignore and impossible to overstate the importance of such a
hypothesis test. Apart from serious numerical instability and potential run-away
effects, possibly caused by improper initialization of or unrealistic
constraints on the optimization procedure, maximum--likelihood inversion will
always return the parameter set \textit{with maximum likelihood}. But whether
the most likely model is, in fact, \textit{any} good, then remains to be
ascertained. Establishing whether eq.~(\ref{Xofk}) in fact holds can be carried
out visually, by inspection of the overlay of the histograms of $2X\ofst$ across
all wave vectors with the probability density function $\chi^2_2$, and by making
`quantile-quantile' plots of the ranked values of $2X\ofst$ versus the inverse
cumulative density function of $\chi^2_2$ evaluated at their corresponding
fractional ranks. Moreover, the two-dimensional map of $X\ofst\ofk$ should show
no residual structure, and will contain information on possible wavenumber
ranges or specific directions (for example, in the presence of anisotropy) in
which the data might be over- or under-fit. All three such representations of
model quality (histograms, quantile-quantile `Q-Q' plots, and two-dimensional
wavenumber maps of the residuals) must be thoroughly scrutinized.

Beyond visual inspection, it is desirable to design a formal test for when the
hypothesis of isotropic Mat\'ern behavior needs to be abandoned, and the
veracity of the parameters recovered by likelihood maximization called into
question, regardless of how narrow their uncertainty intervals based on
eq.~(\ref{magic}) may be. Failing the test could be due, for example, to
non-Gaussianity, or to the presence of patterns or preferred directions
indicating that the data should rather be interpreted under anisotropic
\cite[e.g.,][]{Goff+89b,Herzfeld+99,Olhede2008,Olhede+2014b} extensions of the
model. We save developing alternative hypotheses for future work.

For a given modeled data sample, we propose as a test statistic the
mean-squared deviation from the expected value of the residual ratio,
\begin{equation}\label{stx}
\stx=\norml[X\ofst\ofk-1]^2
.
\end{equation}
The smoothness and boundedness of the spectrum~$\Sbar\ofst\ofk$, the presumed
independence of $X\ofst\ofk$ between wavenumbers, and the central limit theorem
should help the variable~$\stx$ to converge to a normal variate. The central
moments of the $p$th power of chi-squared variables with $m$ degrees of freedom
\cite[]{Davison2003} satisfy $\big\langle\!\left[\chi^2_m\right]^p\!
\big\rangle= 2^p\Gamma(p+m/2)/\Gamma(m/2)$, from which we obtain
\begin{equation}\label{momid}
\big\langle X\ofst^p\big\rangle=\Gamma(p+1)=p!
.
\end{equation}

The case $p=1$ discussed by \cite{Simons+2013} in reducing eq.~(\ref{hessian})
to eq.~(\ref{fisher}), is easily verified. Evaluating eq.~(\ref{momid}) for the
case $p=2$ then yields the expectation of eq.~(\ref{stx}), our test statistic,
\begin{equation}
\langle\stx\rangle=
\norml\big\langle X^2\ofst\ofk-2X\ofst\ofk+1\big\rangle=1
.
\end{equation}
For its variance, again assuming independence between the wave vectors, we find
from elementary calculations that
\begin{align}
MN\var\big\{\stx\big\}&=
\var\big\{X\ofst^2\big\}+4\,\var\big\{X\ofst\big\}-4\,\cov\big\{X\ofst^2,X\ofst\big\}\\
&=\big\langle X\ofst^4 \big\rangle-\big\langle X\ofst^2 \big\rangle^2
+4\big\langle X\ofst^2 \big\rangle-4\big\langle X\ofst \big\rangle^2
-4\big\langle X\ofst^3 \big\rangle+4\big\langle X\ofst^2 \big\rangle\big\langle X\ofst \big\rangle
=8.
\end{align}
Hence we deduce that our chosen metric converges `in law' \cite[]{Ferguson96} to
a variable distributed as:
\begin{equation}\label{inlaw}
\stx\inlaw \mcN(1,8/[MN])
.
\end{equation}
In other words, by computing eq.~(\ref{stx}) after finding the
maximum-likelihood estimates for the Mat\'ern parameters of a data set, we are
in a position to test whether the residuals are distributed according to the
theory, rejecting the model at whichever confidence level we envisage.

\begin{figure*}\centering
\rotatebox{0}{\includegraphics[width=\textwidth,angle=0,trim=0cm 1cm 0cm 1cm,clip]{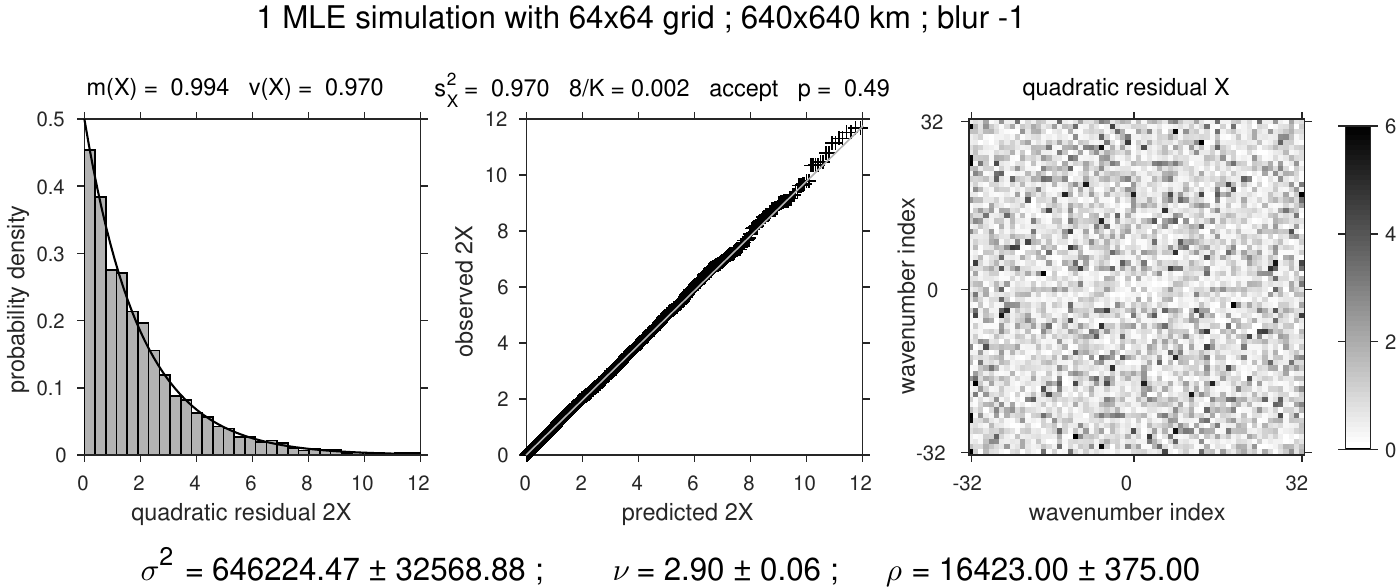}}
\caption{\label{dd}Residual statistics of simulations carried out on a
  101$\times$111 grid, with spacings $\Dx=\Dy=10$~km, convolutional blurring, and
  true Mat\'ern parameter values of $\sigma^2_0=1$~km$^2$, $\nu_0=2.5$,
  $\rho_0=20$~km. Results pertaining to one of the simulations and its
  maximum-likelihood recovery. Distribution of the variable~$X\ofst\ofk$ of
  eq.~(\ref{Xofk}), as a histogram across all wavenumbers with the theoretical
  distribution superposed, as a quantile-quantile plot for the distribution in
  question, and as a spectral-domain map.}
\end{figure*}

Fig.~\ref{dd} enlightens us in this regard. The three panels (histograms, Q-Q
plots, 2-D map) show the result of a successful experiment with parameters
similar to those of the runs presented in Fig.~\ref{mleosl}, in which the
Mat\'ern parameters were very well recovered. The residuals $X\ofst\ofk$ show
the expected distributional behavior without any hint of remaining structure,
privileged directions or otherwise. The sample mean and the sample variance of
the variable~$X\ofst$ are listed above the first panel. Per eq.~(\ref{Xofk}),
both are expected to be one. Above the second panel are the test
statistic~$\stx$, its variance under the null hypothesis, the decision to accept
(in this case) or reject, and the two-sided probability that values more extreme
than the calculated one are likely to occur under the model. All of these
entities factor into our decision making.

\begin{figure*}\centering
\includegraphics[width=0.65\textwidth,angle=0,trim=0cm 0.1cm 10cm 8.1cm,clip]{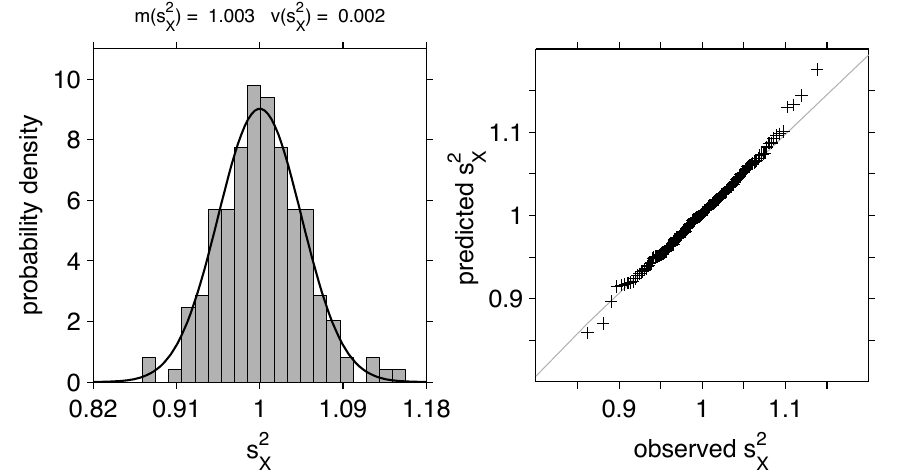}
\caption{\label{momx}The behavior of the test statistic for appropriateness of
  the Mat\'ern model, $\stx$ of eq.~(\ref{stx}), across an ensemble of 175
  simulation and recovery experiments. Histogram and its prediction, and
  quantile-quantile plot comparing observations to predictions. The theoretical
  behavior of eq.~(\ref{inlaw}) is validated.}
\end{figure*}

Fig.~\ref{momx} validates the distribution of the test statistic through a
series of simulation experiments with the same parameters as those used for
Fig.~\ref{dd}. Reporting on the `test of the test', specifically, whether
eq.~(\ref{inlaw}) holds, the two panels illustrate the distribution of $\stx$
across 175 realizations. Sample mean and variance of the test statistic are
labeled above the first panel, which displays the histogram of $\stx$ over the
simulations. The second panel shows the linearity of the Q-Q plots. Our
conclusion is that using $\stx$ as a statistic results in a useful and sensitive
test on the appropriateness of the Mat\'ern model, whatever its parameters, and
irrespectively of their confidence intervals.

\begin{figure*}\centering
  \unboxed{\includegraphics[height=\htwo,angle=-0,trim=1.6cm 16.65cm 1.95cm 1.5cm,clip]{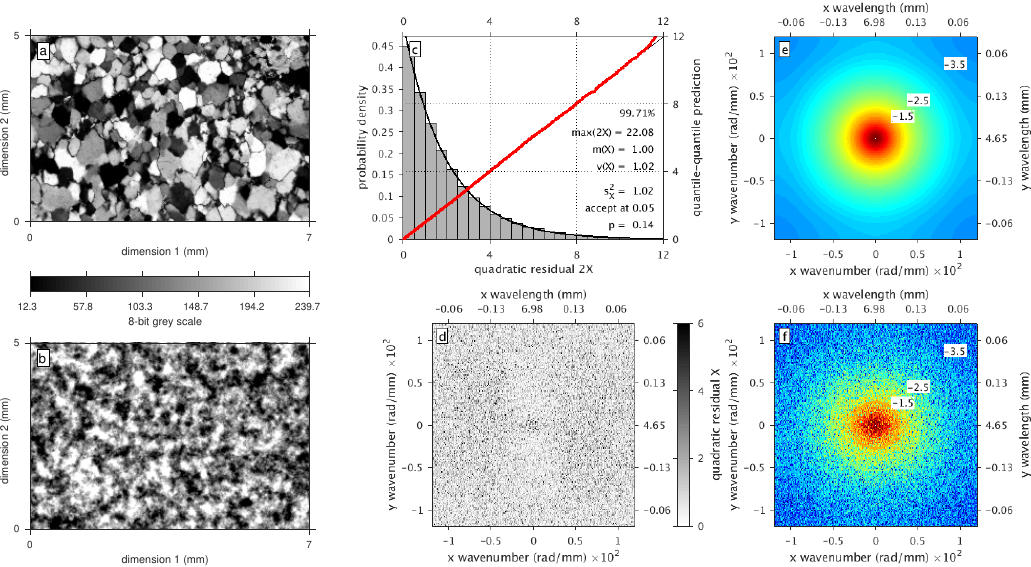}}
  \caption{\label{quartzite} Maximum-likelihood analysis for the Mat\'ern
    covariance structure of a thin section of a quartzite rock under
    cross-polarized light \cite[]{DaMommio+2025}. Shown are the observed data
    (\textit{a}) and a synthetic randomly generated from the Mat\'ern parameters
    recovered as our estimate (\textit{b}) as discussed in Sec.~\ref{fourthree},
    a histogram and a quantile-quantile (Q-Q) plot of the quadratic residual
    $2X\ofst$ (\textit{c}), with $X\ofst\ofk$ being rendered and inspected for
    patterns in wave vector space (\textit{d}), as discussed in
    Sec.~\ref{residuals}. Also shown are the expected (\textit{e}) and
    observed (\textit{f}) periodograms, $\bar{\mcS}\ofst\ofk$ and
    $|H\ofk|^2$ respectively, with contour lines (labeled by their exponent) for
    the former overlain on the latter. The model fits extremely well over the
    entire wavenumber range.}
\end{figure*}

\begin{figure*}\centering
  \unboxed{\includegraphics[height=\htwo,angle=-0,trim=1.6cm 16.8cm 2cm 1.5cm,clip]{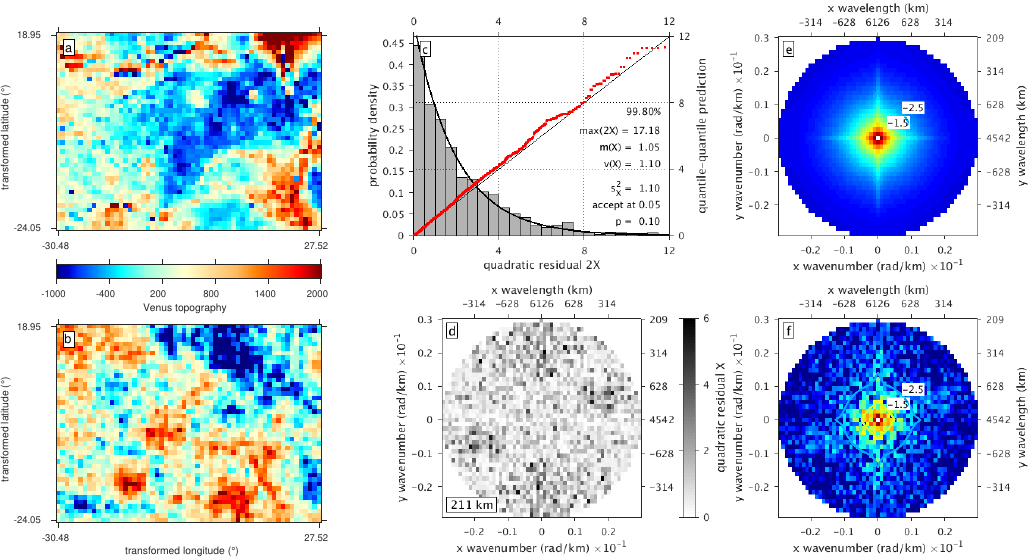}}
  \caption{\label{venus}Maximum-likelihood analysis for the Mat\'ern covariance
    structure of a patch of Venus topography. Layout and annotation are as in
    Fig.~\ref{quartzite}. To counter the isotropic filtering present in the
    original data set \cite[]{Rappaport+99,Eggers2013}, we calculated the
    likelihood and based our decision solely on the wavenumbers within the
    rendered spectral disk. The model fits very well, small residual departures
    visible in the Q-Q plot notwithstanding, which are indeed expected for
    correlated order-statistics at both ends \cite[]{Davison2003}. We draw
    attention to the numerical annotations in the top middle panel. The
    percentile corresponding to the horizontal axis limit is shown. Below that,
    maximum, mean, and variance of the residual $X\ofst$ (see eq.~\ref{Xofk}),
    the test statistic $\stx$ (see eq.~\ref{stx}), the decision at the 95\%
    significance level, and the one-sided exceedance probability for the test of
    it being normally distributed (see eq.~\ref{inlaw}).}
\end{figure*}

\begin{figure*}\centering
  \unboxed{\includegraphics[height=\htwo,angle=-0,trim=2.15cm 16.65cm 2.5cm 1.5cm,clip]{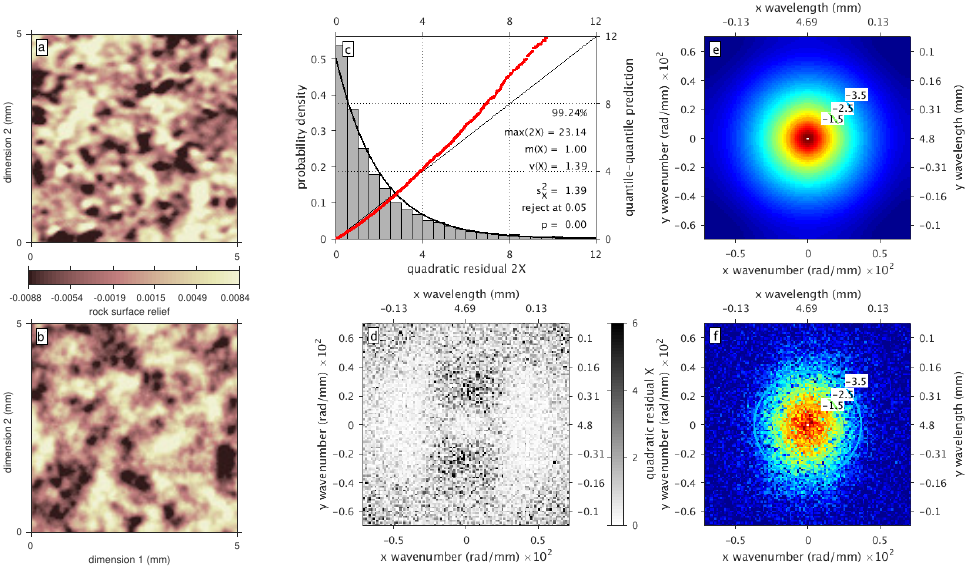}}
  \caption{\label{tribology} Maximum-likelihood analysis for the Mat\'ern
    covariance structure of a saw-cut, polished, and sandblasted granitic rock
    surface fabricated in a friction experiment \cite[]{Guerin+2023}. Layout and
    annotation are as in Figs~\ref{quartzite}--\ref{venus}. The model fits
    relatively poorly under the most intransigent interpretation of the Q-Q
    plot, which could be argued is too demanding, as the `mass' of the
    distribution is reasonably well aligned with statistical expectation. The
    residuals show relatively strong hints of unmodeled structure with
    directional and wavenumber dependence. On the whole, we consider this to be
    an acceptable model, and in future work we might relax the strict adherence
    to eq.~(\ref{inlaw}) as a make-or-break decision for model acceptance.}
\end{figure*}

\begin{figure*}\centering
  \unboxed{\includegraphics[height=\htwo,angle=-0,trim=1.65cm 16.9cm 1.9cm 1.5cm,clip]{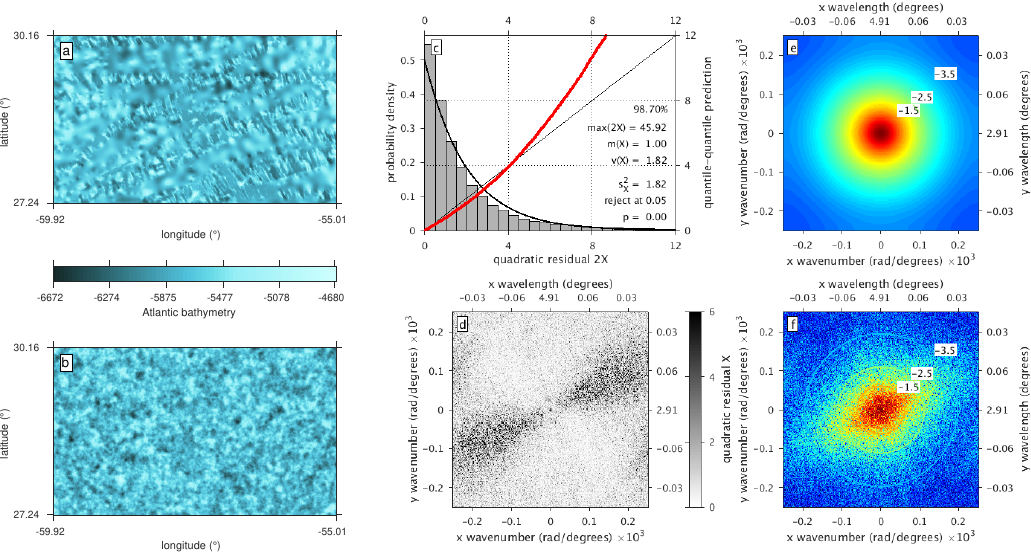}}
  \caption{\label{seafloor} Maximum-likelihood analysis for the Mat\'ern
    covariance structure of Atlantic seafloor bathymetry. Layout and annotation
    as in Figs~\ref{quartzite}--\ref{tribology}. Of all the examples shown so
    far, this model fits most poorly, while it clearly still has much
    merit. However, its failure to be accepted based on the stringent
    eq.~(\ref{inlaw}) will drive us to consider models beyond the isotropic
    stationary Mat\'ern class. The strong directional dependence of the
    residuals reveals that anisotropic behavior will need to be considered, as
    is indeed expected both based on the intrinsic geophysical and geological
    properties of the seafloor \cite[]{Goff+89b}, and from the heterogeneous and
    anisotropic model \cite[]{GEBCO2019} built from compiled multiresolution
    data sources that include both direct shipboard seafloor sounding and
    interpreted satellite altimetry sea-surface measurements \cite[]{Smith+97}.}
\end{figure*}

\section{M~A~X~I~M~U~M~-~L~I~K~E~L~I~H~O~O~D{\hsps}P~R~A~C~T~I~C~E}
\label{modeltest}

Our fundamental point of departure is that every geological or geophysical
spatial data set is a sample, a realization, of a parameterizable random
process, whose parameters we seek to recover. Whether these statistical
parameters then are directly interpretable
\cite[e.g.,][]{Malinverno91,Whittaker+2008}, used for interpolation or
extrapolation \cite[e.g.,][]{Mareschal89,Sandwell+2022}, anomaly detection and
classification \cite[e.g.,][]{Herzfeld+96,Herzfeld+2001}, or in order to infer
from them any kind of other, e.g., geophysical, oceanographic, or geological,
information
\cite[e.g.,][]{Stephenson84,Goff+2010,Song+2003,Persson2006,Sagy+2007,Bottero+2020,Cristini+2012,Grainger+2021,Wunsch2022}
is beyond the immediate remit of this paper.

There may not appear to be a dearth of approaches to solve problems of the
nature described above, but in remote-sensing, geology, geophysics,
oceanography, and planetary science, many methods in common use yield results
that are biased, too variable, or too vague, to be of quantitative statistical
use. An example to help clarify the needs addressed in this paper is how best to
answer the seemingly straightforward question
\cite[e.g.,][]{Munk55,Kreslavsky+2000,Shepard+2001,Grohmann+2010,Candela+2012,Reich+2013,Guerin+2023}:
``how \textit{rough} is that (topographic, ocean-floor, sea-surface,
earthquake-fault, rock-sample, ... tooth) surface?''

The overarching accomplishment of this paper is that we have developed a fast,
efficient, and effective strategy for such parameter estimation problems, for
Gaussian processes governed by a flexible class of two-point covariance
structures, under realistic sampling scenarios which require explicit
consideration and treatment in order to render the solutions interpretable,
intercomparable, and robust, with quantifiable uncertainties and a test for
model appropriateness. The sampling patterns discussed so far, and in the
remainder of this paper, are regularly spaced Cartesian rectangular grids.
Should the boundaries of a region be irregular, or the sampling be incomplete,
as long as the selection function or window $w\ofx$ in eq.~(\ref{fourierHw}) can
accommodate picking out what is actually being observed, our theory and methods
will continue to hold. We will return to incomplete, be it random,
deterministic, or otherwise structured sampling patterns in future work.

The statistical literature abounds with methods and examples of asymptotic
(infill, growing, and mixed-domain) convergence results for specific special
cases of the Mat\'ern spectral density or covariance, that is, for specific
values of~$\nu$ in eqs~(\ref{matern2}) and~(\ref{Kr}), which resolve to simpler
forms \cite[]{Guttorp+2006}. Sec.~\ref{fourthree} showed how the statistical
behavior of the solutions depends both on the sampling and on the random process
being sampled, and it therefore must remain beyond the scope of this paper to
discuss this interplay with any aspiration to generality.  The important cases
of anisotropy, non-stationarity, non-Gaussianity, and correlations between
multivariable fields will await further treatment under our framework. Any and
all of those considerations, however, will rely on a thorough understanding of
the null-hypothesis of univariate, stationary, isotropic, Gaussian fields such
as are provided here.

For a first confrontation of our new results with geophysical and geological
practice, we will limit ourselves to showing a few illuminating examples of
parameter recovery, uncertainty quantification, and model testing, on completely
sampled, rectangular grids. In Figs~\ref{quartzite}--\ref{seafloor}, the left
column contains an image of the data, underneath which we plot a simulation
based on the Mat\'ern model with the parameters derived from our procedure. The
middle column appears in a layout that is a combination of the three panels of
Fig.~\ref{dd}, showing the histogram of
$X\ofst\ofk=\Sbar\ofst^{-1}\hsomm\ofk\hsom|H\ofk|^2$ of eq.~(\ref{Xofk}), its
theoretical probability density (using the left ordinate) and Q-Q plot (using
the right ordinate), the test statistic $\stx$ of eq.~(\ref{stx}) and the
decision based on it via eq.~(\ref{inlaw}). In the panel below we show the
values of $X\ofst\ofk$ in wave vector space, to inspect for any undesirable
patterns that may persist. The panels in the rightmost column are as the
rightmost top panels in Fig.~\ref{square}, showing the blurred spectral density
$\Sbar\ofst\ofk$, that is, the expectation of the windowed periodogram,
$\langle|H\ofk|^2\rangle$, and the periodogram~$|H\ofk|^2$ observed from the
data. Contours of the expected spectrum are drawn on the observed, highlighting
where they match.

The optimization was carried out by a simple unconstrained finite-difference
gradient-based method (using MATLAB's \texttt{fminunc}). Starting with the
sample variance as a candidate for~$\st$, our runs were initialized with values
that, while inspired by comments made by \cite{Vanmarcke83} relative to $\rho$,
and \cite{Sykulski+2019} relative to $\nu$, have not needed to be more than
``reasonable'' guesses. Up to a planar trend was removed for the data prior to
analysis, and a cosine-squared window tapering 10\% all around was applied to
smooth any edges, exactly as we had done to make Figs~\ref{dd}--\ref{momx}. See
Table~\ref{thetable} for a summary of the results discussed next.

The first example is a photomicrograph of a rock sample in a 30~$\mu$m
thin-section used for optical petrography. Fig.~\ref{quartzite} is a gray-scale
image of a quartzite, a metamorphic rock composed almost entirely of the mineral
quartz (SiO$_2$). The field of view is 7~mm, and the image was taken under
cross-polarized light \cite[]{DaMommio+2025}, revealing individual grains with
characteristic undulose extinction. The Mat\'ern model fits very well in this
case, despite the sand grains being distinct entities with relatively constant
grayscale values and visible grain boundaries. The synthetic will not be
mistaken for an actual, possibly non-Gaussian, rock sample, yet it remains
statistically indistinguishable from the perspective of Gaussian field
modeling. Virtually no discernible patterns characterize the spectral-domain
residuals, and the power-spectral densities are almost entirely isotropic, as
expected for this type of uniformly recrystallized metamorphic rock.

The second example is a patch of topography from the planet
Venus. Fig.~\ref{venus} shows the region around Hinemoa Planitia obtained from
radar altimetry mapping by the Magellan spacecraft \cite[]{Rappaport+99},
expanded from spherical harmonics, reprojected to center on $(261.57^\circ,
5.29^\circ)$, and downsampled from the original full resolution
\cite[]{Eggers2013}. The area shown occupies a mean elevation of 150~m above
the mean planetary radius, with a standard deviation of 756~m. The rendered
colors span the 1st through the 99th percentile, which cover relative elevations
1161~m below and 2176~m above the deplaned surface. To accommodate an isotropic
Nyquist filtering step applied to the original data, which effectively erased
all power from the corners outside the circle inscribed in the Cartesian wave
vector rectangle, we restricted the likelihood calculation to the disk-shaped
domain of radius 211~km shown in the spectral plots. The Mat\'ern model fits the
data well and is formally accepted as a null-hypothesis. Nevertheless, there are
visible hints of systematic departures from complete isotropy in the middle- to
large-wavenumber range that may well prove to be geologically significant upon
further analysis. Furthermore, by our definition \cite[]{Eggers2013}, Hinemoa
Planitia as a coherent geological unit is an irregularly bounded subdomain of
the rectangular patch shown, which will require applying spatially selective
windowing to fully and uniquely characterize as a locally stationary random
field.

The third example is a portion of a granite surface taken before a tribological
experiment \cite[]{Guerin+2023}. Fig.~\ref{tribology} shows a piece of the
relief on La Peyratte granite ``sample~R1~top'', measured using a Keyence
VR-3200 surface scanning light microscope, with a reported root-mean-squared 
``asperity'' height of 12--16~$\mu$m.  The ``rough'' surface, created by
sandblasting the saw-cut, pre-polished, sample surface with silicon carbide
(SiC) powder is marked by long-wavelength relief and high gradients between
extreme peaks and troughs which proved challenging to capture in the modeled
field. Hence we deplaned the image and winsorized it within the 7.5 and 92.5
percentiles prior to analysis. The results, while we deem the synthetic visually
encouraging, nevertheless warrant rejection of the Mat\'ern model. The
spectral-domain residuals fail to meet our strict distributional criterion, and
show systematic deviation from what should be a complete lack of structure in
wave vector space. Despite these misgivings, the best-fit solutions for the
Mat\'ern parameters made on successive measurements of rock surface relief as it
gets ground down and smoothed in fault-friction experiments hold promise for
their objective if imperfect characterization.

The fourth and last example involves a portion of the seafloor in the North
Atlantic. Fig.~\ref{seafloor} shows the \cite{GEBCO2019} bathymetric model for
the area, which lies at a mean elevation of 5738~m below the sea surface, with a
standard deviation of 271~m. The rendered colors span the depths between 6672~m
and 4680~m. Oceanic bathymetry is a \textit{model} based on few direct
observations: swaths of high-resolution shipboard-derived multibeam sonar
data draped over a long-wavelength low-resolution model derived from
radar altimetry of the sea surface, which gravitationally reflects the topography
of the ocean bottom \cite[]{Smith+97}. Whether completely sampled or not, ocean
bathymetry, as a mirror of plate tectonics, rarely if ever shows isotropic
behavior. The isotropic Mat\'ern model developed in this paper clearly does not
apply. We include this example because our analysis allows for the testing
(and rejection, as in this case) of isotropic behavior, and the initial
evaluation of anisotropy, for which suitable models and estimation methods need
to be developed.

\begin{table}
    \centering
    \begin{tabular}{rcrrrrrrrrrrrrr}
         Example   & Figure & $K$    &    $\st/s^2$  & $\pm$\% &  $\nu$  & $\pm$\%  & $\rho$ & $\pi\rho/r$ & $\pm$\%  & $\{\st,\nu\}$  & $\{\st,\rho\}$   & $\{\nu,\rho\}$ \\\hline
         Seafloor  & \ref{seafloor}  &   92196  &  0.92  &       3 &  1.26  &  0.6 & 1.75~km  & 0.01  &  2 & $-38$ & $+90$ & $-72$ \\
         Quartzite & \ref{quartzite} &   47526  &  0.91  &       4 &  0.91  &  1.0 & 0.0512~mm & 0.02  &  3 & $-31$ & $+90$ & $-67$ \\
         Granite   & \ref{tribology} &   11556  &  1.64  &      10 &  1.64  &  1.4 & 0.1033~mm & 0.05  &  4 & $-39$ & $+92$ & $-71$ \\
         Venus     & \ref{venus}     &    2026  &  1.64  &      53 &  0.31  &  9.0 & 841~km  & 0.34  & 96 & $-22$ & $+96$ & $-46$ 
    \end{tabular}
    \caption{\label{thetable}Results from our experiments with geological and
      geophysical data, sorted by the degrees of freedom of the analysis, the
      size of the field, $K=M\times N$, adjusted to the wavenumber disk for the
      case of Venus. The estimate for the variance $\st$ is quoted as a fraction
      of the sample variance of the data set, here denoted $s^2$. In the
      presence of significant range, $s^2$ is small relative to $\st$. The
      estimate of the range parameter~$\rho$ is first presented in units of the
      field and then is multiplied by $\pi$, see Fig.~\ref{sdfcovetc}, and
      expressed as a fraction of $r$, the length of the diagonal of the data
      grid.  For all three parameters, the one-standard-deviation estimation
      uncertainty is listed to the nearest per cent of the parameter itself. The
      final three columns contain the correlation between parameter estimates,
      as in Fig.~\ref{cc}, in per cent.}
\end{table}

\section{C~O~N~C~L~U~S~I~O~N~S}

The \cite{Matern60} covariance provides a parsimonious, three-parameter,
characterization of stationary and isotropic geological and geophysical
univariate Gaussian fields. While it has enjoyed choice application over the
years, no truly pragmatic framework existed for the analysis of finite, sampled
spatial data sets, including those with irregular boundaries or missing grid
points. To recover the diagnostic Mat\'ern parameters of variance, smoothness,
and range, exact space-based likelihood methods are almost always impractical or
computationally out of reach. Spectral-domain likelihood approximations
introduced by \cite{Whittle54} hold the advantage of computational and
statistical efficiency, but their parameter estimation bias has been an ongoing
cause for concern, especially for incompletely and irregularly bounded sampled
fields. The debiased spatial Whittle likelihood of \cite{Guillaumin+2022}
correctly accounts for edge effects and data omission by blurring the spectral
density function in a manner that is both exact and fast. We showed its efficacy
in recovering Mat\'ern covariance parameters on synthetically generated and
actually observed geophysical fields. A quasi- or pseudo-likelihood, the
debiased Whittle likelihood is indifferent to wave vector correlations during
parameter estimation. We showed how to calculate these interactions, exactly,
and how to include them to derive the full parameter estimation covariance
matrix. Only when these are properly acknowledged do we produce reliable
uncertainty estimates, as we validated experimentally. Our final addition to
complete the theory involves a test for the epistemic appropriateness of the
Mat\'ern model, against which possible non-stationary, an-isotropic,
non-Gaussian departures from the null-hypothesis can be evaluated.

Our work allows us to ask and confidently answer questions related to the
statistical description of geological and geophysical fields and processes in a
manner that produces solutions that are interpretable, intercomparable, and
robust, with quantifiable uncertainties and tests for overall model
fitness. What numbers capture the essence of a patch of spatial data? Can we
estimate them, can we derive uncertainty bounds for them, can we simulate
``new'' realizations of fields that behave exactly as if they were derived from
similar processes?  The answer to all of these questions is ``yes''.  Our
approach holds in one, two, and three dimensions, and generalizes to multiple
variables, e.g., when topography and gravity are being considered jointly
\cite[e.g., linked by flexure, erosion, or other surface and sub-surface
  modifying processes, see][]{Simons+2013}, or for remote-sensing applications.
While currently agnostic as to the application domain, our ultimate goal is to
derive geophysical ``process'' from statistical ``parameters'', e.g., to be able
to assign likely formation mechanisms and histories for the data under
consideration, with specific input from domain knowledge.

\section{D~A~T~A{\hsps}A~V~A~I~L~A~B~I~L~I~T~Y}
\label{code}

All the code used to conduct the calculations and produce the figures in this
paper made in \textsc{Matlab} by the authors is openly available as
Release~2.0.0 from \url{https://github.com/csdms-contrib/slepian_juliet}, doi:
\texttt{10.5281/zenodo.4085253}. For an alternative code base in Python, see
also \cite{Guillaumin+2026}. Also see the \textit{R} package by
\cite{Paciorek2007}. The quartzite photomicrograph is described by
\cite[]{DaMommio+2025} and was obtained from
\url{https://www.alexstrekeisen.it/english/meta/quartzite.php}, and used with
permission from Dr.~Alessandro Da Mommio (\texttt{alexdm83@libero.it}).  The
original Venus topography model \texttt{shtjv360} was constructed by
\cite{Rappaport+99} and resides at
\url{https://pds-geosciences.wustl.edu/missions/magellan/shadr_topo_grav/index.htm}.
The granite surface data pertain to the paper by \cite{Guerin+2023} and are
available from \url{https://zenodo.org/records/6411819}.  The seafloor
bathymetry model \texttt{GEBCO\_2019} is available from
\url{https://www.gebco.net/data-products/gridded-bathymetry-data/gebco-2019} as
described by \cite{GEBCO2019}.

\section{A~C~K~N~O~W~L~E~D~G~E~M~E~N~T~S}

This work was sponsored by the National Aeronautics and Space Administration
under grant NNX11AQ45G. OLW gratefully acknowledges financial support from the
Schmidt DataX Fund, grant number 22-008, at Princeton University, made possible
through a major gift from the Schmidt Futures Foundation. FJS thanks the
Institute for Advanced Study for a stimulating research environment during
2023--2024, and KU~Leuven and the Groot Begijnhof for enabling productive Summer
months. We thank Erin O'Neil for helpful comments on the final draft. We are
grateful to the Associate Editor Carl Tape, and to John Goff and Michael Stein
for constructive reviews that improved the manuscript.

\bibliographystyle{gji}
\bibliography{bib.bib}

@string{AGG = {Ann.~Geophys.--Germany}}

@string{AP = {Academic Press}}

@string{APJ = {Astroph.~J.}}

@string{AS = {Ann.~Stat.}}

@string{AnA = {Astron.~Astroph.}}

@string{AT = {Acoustics Today}}

@string{CG = {Comput.~Geosci.}}

@string{CUP = {Cambridge Univ.~Press}}

@string{EAP = {Elsevier Academic Press}}

@string{EJS = {Electron.~J.~Stat.}}

@string{EPSL = {Earth~Planet.~Sci.~Lett.}}

@string{ESS = {Earth Space Sci.}}

@string{GC = {Geochem.~Geophys.~Geosys.}}

@string{GJI = {Geophys.~J.~Int.}}

@string{GJRAS = {Geophys.~J.~R.~Astron.~Soc.}}

@string{GRL = {Geophys.~Res.~Lett.}}

@string{IEEE-GRS = {IEEE T.~Geosci.~Remote}}

@string{IEEE-IT = {IEEE T.~Inform.~Theory}}

@string{IEEE-OE = {IEEE J.~Ocean.~Eng.}}

@string{IEEE-SP = {IEEE T.~Signal~Process.}}

@string{JASA = {J.~Acoust.~Soc.~Am.}}

@string{JASTA = {J.~Am.~Stat.~Assoc.}}

@string{JC = {J.~Climate}}

@string{JCGS = {J.~Comput.~Graph.~Stat.}}

@string{JGR = {J.~Geophys.~Res.}}

@string{JMR = {J.~Mar.~Res.}}

@string{JOSS = {J.~Open Source Softw.}}

@string{JRSSB = {J.~R.~Stat.~Soc., Ser.~B}}

@string{JSPI = {J.~Stat.~Plann.~Infer.}}

@string{JTSA = {J.~Time~Ser.~Anal.}}

@string{NPG = {Nonlin.~Proc.~Geophys.}}

@string{OUP = {Oxford Univ.~Press}}

@string{PAGEOPH = {Pure Appl.~Geophys.}}

@string{SERRA = {Stoch.~Env.~Res.\ Risk~A.}}

@string{SPS = {Spat.~Stat.}}

@string{SSC = {Stat.~Sci.}}

@string{SSR = {Surf.~Sci.~Rep.}}

@string{WRR = {Water Resources Res.}}

@Book{Abramowitz+65,
  author =	 {Milton Abramowitz and Irene A. Stegun},
  title =	 {Handbook of Mathematical Functions},
  publisher =	 {Dover},
  year =	 1965,
  address =	 {New York}
}

@Book{Adler1981,
  title =	 {The Geometry of Random Fields},
  author =	 {Robert J. Adler},
  year =	 1981,
  publisher =	 {Wiley},
  address =	 {New York},
}

@Article{Aharonson+98,
  author =	 {Oded Aharonson and Maria T. Zuber and Gregory
                  A. Neumann and James W. Head},
  title =	 {{M}ars: {N}orthern hemisphere slopes and slope
                  distributions},
  journal =	 GRL,
  year =	 1998,
  volume =	 25,
  number =	 24,
  pages =	 {4413--4416}
}

@Article{Appourchaux+98a,
  author =	 {T. Appourchaux and L. Gizon and
                  M.-C. Rabello-Soares},
  title =	 {The art of fitting {p}-mode spectra. {I}. {M}aximum
                  likelihood estimation},
  journal =	 {Astron.~Astroph.~Suppl.~Ser.},
  year =	 1998,
  volume =	 132,
  pages =	 {107--119, doi: 10.1051/aas:1998441}
}

@Article{Appourchaux+98b,
  author =	 {T. Appourchaux and M.-C. Rabello-Soares and
                  L. Gizon},
  title =	 {The art of fitting {p}-mode spectra. {II}. {L}eakage
                  and noise covariance matrices},
  journal =	 {Astron.~Astroph.~Suppl.~Ser.},
  year =	 1998,
  volume =	 132,
  pages =	 {121--132, doi: 10.1051/aas:1998440}
}

@Book{Aster+2018,
  author =	 {Richard C. Aster and Brian Borchers and Clifford
                  H. Thurber},
  title =	 {Parameter Estimation {a}nd Inverse Problems},
  publisher =	 EAP,
  year =	 2018,
  edition =	 3,
  address =	 {San Diego, Calif.}
}

@Article{Baig+2003,
  author =	 {Adam M. Baig and F. A. Dahlen and S.-H. Hung},
  title =	 {Traveltimes of waves in three-dimensional random
                  media},
  journal =	 GJI,
  year =	 2003,
  volume =	 153,
  number =	 2,
  pages =	 {467--482, doi: 10.1046/j.1365-246X.2003.01905.x}
}

@Article{Baig+2004a,
  title =	 {Statistics of traveltimes and amplitudes in random media},
  author =	 {A. M. Baig and F. A. Dahlen},
  journal =	 GJI,
  volume =	 158,
  number =	 1,
  pages =	 {187--210, doi: 10.1111/j.1365-246X.2004.02300.x},
  year =	 2004
}

@Article{Becker+2007b,
  title =	 {Stochastic analysis of shear-wave splitting length
                  scales},
  author =	 {T. W. Becker and J. T. Browaeys and T. H. Jordan},
  journal =	 EPSL,
  volume =	 259,
  number =	 {3--4},
  pages =	 {526--540, doi: 10.1016/j.epsl.2007.05.010},
  year =	 2007
}

@Book{Bendat+2000,
  author =	 "Julius S. Bendat and Allan G. Piersol",
  title =	 "Random Data: {A}nalysis {a}nd Measurement
                  Procedures",
  publisher =	 "John Wiley",
  year =	 2000,
  address =	 "New York",
  edition =	 3
}

@Article{Botev+2010,
  title =	 {Kernel density estimation via diffusion},
  author =	 {Zdravko I. Botev and J. F. Grotowski and
                  D. P. Kroese},
  journal =	 AS,
  volume =	 38,
  number =	 5,
  pages =	 {2916--2957, doi: 10.1214/10-AOS799},
  year =	 2010
}

@article{Bottero+2020,
  title =	 {On the influence of slopes, source, seabed and water column
                  properties on {T} waves: {G}eneration at shore},
  author =	 {Alexis Bottero and Paul Cristini and Dimitri Komatitsch},
  journal =	 PAGEOPH,
  volume =	 177,
  pages =	 {5695--5711, doi: 10.1007/s00024-020-02611-z},
  year =	 2020
}

@Article{Candela+2012,
  author =	 {Thibault Candela and Fran\c{c}ois Renard and Yann
                  Klinger and Karen Mair and Jean Schmittbuhl and
                  Emily E. Brodsky},
  title =	 {Roughness of fault surfaces over nine decades of
                  length scales},
  journal =	 JGR,
  year =	 2012,
  volume =	 117,
  pages =	 {B08409, doi: 10.1029/2011JB009041, 2012}
}

@Article{Carpentier+2007,
  title =	 {Underestimation of scale lengths in stochastic
                  fields and their seismic response: {a}
                  quantification exercise},
  author =	 {S. Carpentier and Kabir Roy-Chowdhury},
  journal =	 GJI,
  volume =	 169,
  number =	 2,
  pages =	 {547--562},
  year =	 2007
}

@Book{Christakos92,
  title =	 {Random Field Models {i}n Earth Sciences},
  author =	 {George Christakos},
  year =	 1992,
  publisher =	 AP,
  address =	 {San Diego, Calif.},
  edition =	 2,
}

@Article{Chu2023,
  author =	 {Tingjin Chu},
  title =	 {Mixed domain asymptotics for geostatistical processes},
  journal =	 {Statistica {S}inica},
  volume =	 33,
  number =	 1,
  year =	 2023,
  pages =	 {551--571, doi: 10.5705/ss.202020.0092}
}

@Book{Cox+74,
  author =	 {D. R. Cox and D. V. Hinkley},
  title =	 {Theoretical Statistics},
  publisher =	 {Chapman and Hall},
  year =	 1974,
  address =	 "London, UK"
}

@Article{Cramer42,
  author =	 {Harald Cram\'er},
  title =	 {On harmonic analysis in certain fuctional spaces},
  journal =	 {Arkiv Mat.~Astr.~Fysik},
  year =	 1942,
  volume =	 {28B},
  pages =	 {1--7}
}

@book{Cressie93,
  author =	 {Noel Cressie},
  publisher =	 {John Wiley},
  title =	 {Statistics for Spatial Data},
  address =	 {London, UK},
  year =	 1993
}

@article{Cristini+2012,
  title =	 {Some illustrative examples of the use of a
                  spectral-element method in ocean acoustics},
  author =	 {Paul Cristini and Dimitri Komatitsch},
  journal =	 JASA,
  volume =	 131,
  number =	 3,
  pages =	 {EL229--EL235, doi: 10.1121/1.3682459},
  year =	 2012
}

@book{DaMommio+2025,
  title =	 {Atlas of Minerals and Igneous and Metamorphic Rocks in
                  Thin-Section},
  author =	 {Alessandro {Da Mommio} and Victoria Pease},
  year =	 2025,
  publisher =	 CUP,
  address =	 {Cambridge, UK}
}

@Article{Dahlhaus+87,
  author =	 {R. Dahlhaus and H. K\"unsch},
  title =	 {Edge effects and efficient parameter estimation for
                  stationary random fields},
  journal =	 {Biometrika},
  volume =	 74,
  number =	 4,
  year =	 1987,
  pages =	 {877--882, doi: 10.1093/biomet/74.4.877}
}

@inproceedings{Dahlhaus1984,
  title =	 {Parameter estimation of stationary processes with spectra
                  containing strong peaks},
  author =	 {Rainer Dahlhaus},
  booktitle =	 {Robust and Nonlinear Time Series Analysis: Proceedings of a
                  Workshop Organized by the Sonderforschungsbereich 123
                  ``Stochastische Mathematische Modelle'', Heidelberg 1983},
  pages =	 {50--67, doi: 10.1007/978\-1\-4615\-7821\-5\_4},
  year =	 1984,
  organization = {Springer}
}

@Book{Davison2003,
  author =	 {A. C. Davison},
  title =	 {Statistical Models},
  publisher =	 CUP,
  year =	 2003,
  address =	 {Cambridge, UK}
}

@article{Deb+2017,
  title =	 {An asymptotic theory for spectral analysis of random fields},
  author =	 {Soudeep Deb and Mohsen Pourahmadi and Wei Biao Wu},
  year =	 2017,
  journal =	 EJS,
  volume =	 11,
  pages =	 {4297--4322, doi: 10.1214/17-EJS1326}
}

@InCollection{Dzhaparidze+83,
  author =	 {K. O. Dzhaparidze and A. M. Yaglom},
  title =	 {Spectrum parameter estimation in time series
                  analysis},
  editor =	 {P. Krishnaiah},
  booktitle =	 {Developments in Statistics},
  pages =	 {1--96, doi: 10.1016/B978-0-12-426604-9.50007-0},
  publisher =	 AP,
  year =	 1983,
  volume =	 4,
  address =	 {New York}
}

@MastersThesis{Eggers2013,
  author =	 {Gabriel Logan Eggers},
  title =	 {A regionalized maximum-likelihood estimation of the
                  spatial structure of {V}enusian Topography},
  school =	 {Princeton University},
  year =	 2013,
  type =	 {{A.~B.~Thesis}}
}

@book{Ferguson96,
  title =	 {A Course in Large Sample Theory},
  author =	 {Thomas S. Ferguson},
  year =	 1996,
  publisher =	 {Chapman and Hall/CRC Press},
  address =	 {New York}
}

@article{Fournier+2014,
  title =	 {Generalization of the noise model for time-distance
                  helioseismology},
  author =	 {Damien Fournier and Laurent Gizon and Thorsten Hohage and
                  Aaron C. Birch},
  journal =	 AnA,
  volume =	 567,
  pages =	 {A137, doi: 10.1051/0004-6361/201423580},
  year =	 2014
}

@Article{Fuentes2007,
  author =	 {Montserrat Fuentes},
  title =	 {Approximate likelihood for large irregularly spaced
                  spatial data},
  journal =	 JASA,
  year =	 2007,
  volume =	 102,
  number =	 477,
  pages =	 {321--331, doi: 10.1198/016214506000000852}
}

@TechReport{GEBCO2019,
  author =	 {{GEBCO~Bathymetric~Compilation~Group}},
  title =	 {The {GEBCO\_2019} grid---{A} continuous terrain model
                  of the global oceans and land},
  institution =	 {British Oceanographic Data Centre, National
                  Oceanography Centre, NERC},
  year =	 2019
}

@article{Geoga+2023,
  title =	 {Fitting {M}at{\'e}rn smoothness parameters using automatic
                  differentiation},
  author =	 {Christopher J. Geoga and Oana Marin and Michel Schanen and
                  Michael L. Stein},
  journal =	 {Stat.~Comput.},
  volume =	 33,
  number =	 2,
  pages =	 {48, doi: 10.1007/s11222-022-10127-w},
  year =	 2023
}

@Article{Gizon+2004,
  title =	 {Time-distance helioseismology: {n}oise estimation},
  author =	 {Laurent Gizon and Aaron C. Birch},
  journal =	 APJ,
  volume =	 614,
  number =	 1,
  pages =	 {472--489, doi: 10.1086/423367},
  year =	 2004
}

@Article{Gneiting+2012,
  author =	 {Tilmann Gneiting and Hana \v{S}ev\v{c}{\'i}kov{\'a}
                  and Donald B. Percival},
  title =	 {Estimators of fractal dimension: {A}sessing the
                  roughness of time series and spatial data},
  journal =	 SSC,
  year =	 2012,
  volume =	 27,
  number =	 2,
  pages =	 {247--277, doi: 10.1214/11-STS370}
}

@Article{Goff+2010,
  title =	 {Global prediction of abyssal hill roughness
                  statistics for use in ocean models from digital maps
                  of paleo-spreading rate, paleo-ridge orientation,
                  and sediment thickness},
  author =	 {John A. Goff and Brian K. Arbic},
  journal =	 {Ocean Modelling},
  volume =	 32,
  number =	 {1--2},
  pages =	 {36--43, doi: 10.1016/j.ocemod.2009.10.001},
  year =	 2010
}

@Article{Goff+88,
  author =	 "John A. Goff and Thomas H. Jordan",
  title =	 "Stochastic modeling of seafloor morphology:
                  {I}nversion of {S}ea {B}eam data for second-order
                  statistics",
  journal =	 JGR,
  year =	 1988,
  volume =	 93,
  number =	 "B11",
  pages =	 "13589--13608, doi: 10.1029/JB093iB11p13589"
}

@Article{Goff+89a,
  author =	 "John A. Goff and Thomas H. Jordan",
  title =	 "Stochastic modeling of seafloor morphology:
                  {R}esolution of topographic parameters by {S}ea
                  {B}eam data",
  journal =	 IEEE-OE,
  year =	 1989,
  volume =	 14,
  number =	 4,
  pages =	 "326--337, doi: 10.1109/48.35983"
}

@Article{Goff+89b,
  author =	 "John A. Goff and Thomas H. Jordan",
  title =	 "Stochastic modeling of seafloor morphology: {A}
                  parameterized {G}aussian model",
  journal =	 GRL,
  year =	 1989,
  volume =	 16,
  number =	 1,
  pages =	 "45--48, doi: 10.1029/GL016i001p00045"
}

@Book{Gradshteyn+2000,
  author =	 {I. S. Gradshteyn and I. M. Ryzhik},
  title =	 {Tables of Integrals, Series, and Products},
  publisher =	 AP,
  year =	 2000,
  address =	 {San Diego, Calif.},
  edition =	 6
}

@article{Grainger+2021,
  title =	 {Estimating the parameters of ocean wave spectra},
  author =	 {Jake P. Grainger and Adam M. Sykulski and Philip Jonathan and
                  Kevin Ewans},
  journal =	 {Ocean.~Eng.},
  volume =	 229,
  pages =	 {108934, doi: 10.1016/j.oceaneng.2021.108934},
  year =	 2021
}

@article{Grainger+2025,
  title =	 {Spectral estimation for spatial point processes and random
                  fields},
  author =	 {Jake P. Grainger and Tuomas A. Rajala and David J. Murrell and
                  Sofia Olhede C.},
  journal =	 {Biometrika},
  pages =	 {asaf089, 10.1093/biomet/asaf089},
  year =	 2025
}

@Article{Grohmann+2010,
  author =	 {Carlos Henrique Grohmann and Mike J. Smith and Claudio
                  Riccomini},
  title =	 {Multiscale analysis of topographic surface roughness in
                  the {M}idland {V}alley, {S}cotland},
  journal =	 IEEE-GRS,
  year =	 2011,
  volume =	 49,
  number =	 4,
  pages =	 {1200--1213, doi: 10.1109/TGRS.2010.2053546}
}

@Article{Gudmundsson+90,
  title =	 {Stochastic analysis of global traveltime data:
                  {m}antle heterogeneity and random errors in the {ISC}
                  data},
  author =	 {{\'O}lafur Gu{$\eth$}mundsson and J. H. Davies and
                  R. W. Clayton},
  journal =	 GJI,
  volume =	 102,
  number =	 1,
  pages =	 {25--43, doi: 10.1111/j.1365-246X.1990.tb00528.x},
  year =	 1990
}

@Article{Guerin+2023,
  title =	 {Preparatory slip in laboratory faults: {E}ffects of roughness
                  and load point velocity},
  author =	 {Simon Gu{\'e}rin-Marthe and Grzegorz Kwiatek and Lei Wang and
                  Audrey Bonnelye and Patricia Mart{\'\i}nez-Garz{\'o}n and
                  Georg Dresen},
  journal =	 JGR,
  volume =	 128,
  number =	 4,
  pages =	 {e2022JB025511, doi: 10.1029/2022JB025511},
  year =	 2023
}

@Article{Guillaumin+2017,
  title =	 {Analysis of non-stationary modulated time series with
                  applications to oceanographic surface flow measurements},
  author =	 {A. P. Guillaumin and A. M. Sykulski and S. C. Olhede and
                  J. J. Early and J. M. Lilly},
  journal =	 JTSA,
  volume =	 38,
  number =	 5,
  pages =	 {668--710, doi: 10.1111/jtsa.12244},
  year =	 2017
}

@Article{Guillaumin+2022,
  author =	 {Arthur P. Guillaumin and Adam M. Sykulski and Sofia
                  Charlotta Olhede and Frederik J. Simons},
  title =	 {The debiased spatial {W}hittle likelihood},
  journal =	 JRSSB,
  volume =	 84,
  number =	 4,
  year =	 2022,
  pages =	 {1526--1557, doi: 10.1111/rssb.12539}
}

@Article{Guillaumin+2026,
  author =	 {Arthur P. Guillaumin and Thomas Goodwin and Olivia L. Walbert,
                  Adam M. Sykulski and Sofia C.Olhede and Frederik J. Simons},
  title =	 {{DSWL} package: {a} {P}ython implementation of the {D}ebiased
                  {S}patial {W}hittle {L}ikelihood},
  journal =	 JOSS,
  volume =	 {X},
  number =	 {Y},
  year =	 2026,
  pages =	 {in revision}
}

@article{Guinness+2017,
  title =        {Circulant embedding of approximate covariances for
                  inference from {G}aussian data on large lattices},
  author =       {J. Guinness and M. Fuentes},
  journal =      JCGS,
  volume =       26,
  number =       1,
  pages =        {88--97, doi: 10.1080/10618600.2016.1164534},
  year =         2017
}

@article{Guinness2019,
  title =	 {Spectral density estimation for random fields via periodic
                  embeddings},
  author =	 {Joseph Guinness},
  journal =	 {Biometrika},
  year =	 2019,
  volume =	 106,
  number =	 2,
  pages =	 {267--286, doi: 10.1093/biomet/asz004}
}

@article{Guttorp+2006,
  title =	 {Studies in the history of probability and statistics
                  {XLIX}. {O}n the {M}at\'ern correlation family},
  author =	 {Peter Guttorp and Tilmann Gneiting},
  year =	 2006,
  pages =	 {989--995, doi: 10.1093/biomet/93.4.989},
  volume =	 93,
  number =	 4,
  journal =	 {Biometrika}
}

@article{Guyon82,
  author =	 {Xavier Guyon},
  title =	 {Parameter estimation for a stationary process on a
                  $d$-dimensional lattice},
  volume =	 69,
  number =	 1,
  pages =	 {95--105, doi: 10.1093/biomet/69.1.95},
  year =	 1982,
  journal =	 {Biometrika}
}

@InCollection{Hamilton2009a,
  title =	 {Power Spectrum Estimation. {I.} {B}asics},
  author =	 {A. J. S. Hamilton},
  booktitle =	 {Data Analysis in Cosmology},
  pages =	 {415--431, doi: 10.1007/978-3-540-44767-2\_12},
  year =	 2009,
  publisher =	 {Springer},
  editor =	 {V. Martinez and E. Saar and E. Gonzales and M. Pons-Borderia},
  volume =	 665,
  series =	 {Lecture Notes in Physics},
  address =	 {Berlin, Germany}
}

@InCollection{Hamilton2009b,
  title =	 {Power spectrum estimation {II.} {L}inear maximum likelihood},
  author =	 {A. J. S. Hamilton},
  booktitle =	 {Data Analysis in Cosmology},
  pages =	 {433--456, doi: 10.1007/978-3-540-44767-2\_13},
  year =	 2009,
  publisher =	 {Springer},
  editor =	 {V. Martinez and E. Saar and E. Gonzales and M. Pons-Borderia},
  volume =	 665,
  series =	 {Lecture Notes in Physics},
  address =	 {Berlin, Germany}
}

@Article{Handcock+93,
  author =	 {Mark S. Handcock and Michael L. Stein},
  title =	 {A {B}ayesian analysis of kriging},
  journal =	 {Technom.},
  year =	 1993,
  volume =	 35,
  number =	 4,
  pages =	 {403--410, doi: 10.1080/00401706.1993.10485354}
}

@Article{Handcock+94,
  author =	 {Mark S. Handcock and James R. Wallis},
  title =	 {An approach to statistical spatial-temporal modeling
                  of meteorological fields},
  journal =	 JASA,
  year =	 1994,
  volume =	 89,
  number =	 426,
  pages =	 {368--378}
}

@Article{Hastings1970,
  author =	 {W. K. Hastings},
  title =	 {Monte {C}arlo sampling methods using {M}arkov chains
                  and their applications},
  journal =	 {Biometrika},
  year =	 1970,
  volume =	 57,
  number =	 1,
  pages =	 {97--108, doi: 10.1093/biomet/57.1.97}
}

@Article{Herzfeld+2001,
  author =	 {Ute Herzfeld and Oliver Zahner},
  title =	 {A connectionist-geostatistical approach to automated
                  image classification, applied to the analysis of
                  crevasse patterns in surging ice},
  journal =	 CG,
  year =	 2001,
  volume =	 27,
  pages =	 {499--512, doi: 10.1016/S0098-3004(00)00089-3}
}

@Article{Herzfeld+95,
  author =	 {Ute C. Herzfeld and Isaac I. Kim and John A. Orcutt},
  title =	 {Is the ocean floor a fractal?},
  journal =	 {Math.~Geol.},
  year =	 1995,
  volume =	 27,
  number =	 3,
  pages =	 {421--462, doi: 10.1007/BF02084611}
}

@Article{Herzfeld+96,
  author =	 {Ute C. Herzfeld and Chris A. Higginson},
  title =	 {Automated geostatistical seafloor classification --—
                  {P}rinciples, parameters, feature vectors, and
                  discrimination criteria},
  journal =	 CG,
  year =	 1996,
  volume =	 22,
  number =	 1,
  pages =	 {35--52, doi: 10.1016/0098-3004(96)89522-7}
}

@Article{Herzfeld+99,
  author =	 {Ute C. Herzfeld and Christoph Overbeck},
  title =	 {Analysis and simulation of scale-dependent fractal
                  surfaces with application to seafloor morphology},
  journal =	 CG,
  year =	 1999,
  volume =	 25,
  number =	 1,
  pages =	 {979--1007, doi: 10.1016/S0098-3004(99)00062-X}
}

@Book{Hofmann+2006,
  title =	 {Physical Geodesy},
  author =	 {Bernhard Hofmann-Wellenhof and Helmut Moritz},
  year =	 2006,
  publisher =	 {Springer},
  edition =	 2,
  address =	 "New York"
}

@article{Hosoya+1982,
  title =	 {A central limit theorem for stationary processes and the
                  parameter estimation of linear processes},
  author =	 {Yuzo Hosoya and Masanobu Taniguchi},
  journal =	 AS,
  pages =	 {132--153, doi: 10.1007/978\-1\-4615\-7821\-5\_4},
  year =	 1982
}

@Article{Isserlis18,
  author =	 {Leon Isserlis},
  title =	 {On a formula for the product-moment coefficient of
                  any order of a normal frequency distribution in any
                  number of variables},
  journal =	 {Biometrika},
  year =	 1918,
  volume =	 12,
  number =	 {1--2},
  pages =	 {134--139, doi: 10.2307/2331932}
}

@Book{Journel+78,
  author =	 "A. G. Journel and C. J. Huijbregts",
  title =	 "Mining Geostatistics",
  publisher =	 AP,
  year =	 1978,
  address =	 "San Diego, Calif."
}

@article{Katzfuss+2021,
  title =	 {A general framework for {V}ecchia approximations of {G}aussian
                  processes},
  author =	 {Matthias Katzfuss and Joseph Guinness},
  journal =	 SSC,
  volume =	 36,
  number =	 1,
  pages =	 {124--141, doi: 10.1214/19-STS755},
  year =	 2021
}

@Article{Kent+96,
  author =	 {John T. Kent and Kanti V. Mardia},
  title =	 {Spectral and circulant approximations to the
                  likelihood for stationary {G}aussian random fields},
  journal =	 JSPI,
  year =	 1996,
  volume =	 50,
  pages =	 {379--394, doi: 10.1016/0378-3758(95)00065-8}
}

@Article{Kitanidis+85,
  title =	 {Maximum likelihood parameter estimation of
                  hydrologic spatial processes by the {G}auss-{N}ewton
                  method},
  author =	 {Peter K. Kitanidis and Robert W. Lane},
  journal =	 {J.~Hydrol.},
  volume =	 79,
  number =	 1,
  pages =	 {53--71, doi: 10.1016/0022-1694(85)90181-7},
  year =	 1985
}

@Article{Kreslavsky+2000,
  author =	 {Mikhail A. Kreslavsky and James W. Head},
  title =	 {Kilometer-scale roughness of {M}ars: {R}esults from
                  {MOLA} data analysis},
  journal =	 JGR,
  volume =	 105,
  number =	 {E11},
  pages =	 {26695--26711, doi: 10.1029/2000JE001259},
  year =	 2000,
}

@InCollection{Kroese+2015,
  author =	 {Dirk P. Kroese and Zdravko I. Botev},
  title =	 {Spatial Process Simulation},
  booktitle =	 {Stochastic geometry, spatial statistics and random fields},
  pages =	 {369--404, doi: 10.1007/978-3-319-10064-7\_12},
  publisher =	 {Springer},
  year =	 2015,
  editor =	 {Volker Schmidt},
  chapter =	 12,
  address =	 {Heidelberg, Germany}
}

@Book{Lambeck88,
  author =	 "Kurt Lambeck",
  title =	 "Geophysical Geodesy",
  publisher =	 OUP,
  year =	 1988,
  address =	 "New York"
}

@article{Landais+2019,
  title =	 {Multifractal topography of several planetary bodies in the
                  solar system},
  author =	 {Fran{\c{c}}ois Landais and Fr{\'e}d{\'e}ric Schmidt and Shaun
                  Lovejoy},
  journal =	 {Icarus},
  volume =	 319,
  pages =	 {14--20, doi: 10.1016/j.icarus.2018.07.005},
  year =	 2019
}

@article{Lilly+2017,
  title =	 {Fractional {B}rownian motion, the {M}at{\'e}rn process, and
                  stochastic modeling of turbulent dispersion},
  author =	 {Jonathan M. Lilly and Adam M. Sykulski and Jeffrey J. Early
                  and Sofia C. Olhede},
  journal =	 NPG,
  volume =	 24,
  number =	 3,
  pages =	 {481--514, doi: 10.5194/npg-24-481-2017},
  year =	 2017
}

@article{Lindgren+2011,
  title =	 {An explicit link between {G}aussian fields and {G}aussian
                  {M}arkov random fields: {t}he stochastic partial differential
                  equation approach},
  author =	 {Finn Lindgren and H{\aa}vard Rue and Johan Lindstr{\"o}m},
  journal =	 JRSSB,
  volume =	 73,
  number =	 4,
  pages =	 {423--498, doi: 10.1111/j.1467-9868.2011.00777.x},
  year =	 2011
}

@Article{Malinverno91,
  author =	 {Alberto Malinverno},
  title =	 {Inverse square-root dependence of mid-ocean-ridge
                  flank roughness on spreading rate},
  journal =	 {Nature},
  year =	 1991,
  volume =	 352,
  pages =	 {58--60, doi: 10.1038/352058a0}
}

@Article{Mardia+84,
  author =	 {Kanti V. Mardia and R. J. Marshall},
  title =	 {Maximum likelihood estimation of models for residual
                  covariance in spatial regression},
  journal =	 {Biometrika},
  year =	 1984,
  volume =	 71,
  number =	 1,
  pages =	 {135--146, doi: 10.2307/2336405}
}

@Article{Mareschal89,
  author =	 {Jean-Claude Mareschal},
  title =	 {Fractal reconstruction of sea-floor topography},
  journal =	 PAGEOPH,
  year =	 1989,
  volume =	 131,
  number =	 {1--2},
  pages =	 {197--210}
}

@Book{Matern60,
  author =	 {Bertil Mat\'ern},
  title =	 {Spatial Variation. Stochastic models {a}nd their application {t}o
                  some problems {i}n forest surveys and other sampling
                  investigations},
  publisher =	 {Statens Skogsforskningsintitut},
  year =	 1960,
  volume =	 49,
  issue =	 5,
  address =	 {Stockholm, Sweden}
}

@article{Metropolis+1949,
  title =	 {The {M}onte {C}arlo method},
  author =	 {Nicholas Metropolis and Stanislaw Ulam},
  journal =	 JASTA,
  volume =	 44,
  number =	 247,
  pages =	 {335--341, doi: 10.1080/01621459.1949.10483310},
  year =	 1949
}

@Article{Montagner86,
  author =	 {Jean-Paul Montagner},
  title =	 {Regional three-dimensional structures using
                  long-period surface waves},
  journal =	 AGG,
  year =	 1986,
  volume =	 4,
  number =	 {B3},
  pages =	 {283--294}
}

@article{Munk55,
  title =	 {High frequency spectrum of ocean waves},
  author =	 {Walter H. Munk},
  journal =	 JMR,
  volume =	 14,
  number =	 4,
  pages =	 {302--314},
  year =	 1955
}

@Article{Neeser+93,
  author =	 {Fredy D. Neeser and James L. Massey},
  title =	 {Proper complex random processes with applications to
                  information theory},
  journal =	 IEEE-IT,
  year =	 1993,
  volume =	 39,
  number =	 4,
  pages =	 {1293--1302}
}

@Article{North+2011,
  author =	 {Gerald R. North and Jue Wang and Marc G. Genton},
  title =	 {Correlation models for temperature fields},
  journal =	 JC,
  year =	 2011,
  volume =	 24,
  pages =	 {5850--5862, doi: 10.1175/2011JCLI4199.1}
}

@Article{Olhede+2014b,
  author =	 {Sofia C. Olhede and David Ram{\'i}res and Peter
                  J. Schreier},
  title =	 {Detecting directionality in random fields using the
                  monogenic signal},
  journal =	 IEEE-IT,
  year =	 2014,
  volume =	 60,
  number =	 10,
  pages =	 {6491--6510, doi: 10.1109/TIT.2014.2342734}
}

@article{Olhede2008,
  title =	 {Localisation of geometric anisotropy},
  author =	 {Sofia C. Olhede},
  journal =	 IEEE-SP,
  volume =	 56,
  number =	 5,
  pages =	 {2133--2138, doi: 10.1109/TSP.2007.912894},
  year =	 2008
}

@Article{Paciorek+2006,
  title =	 {Spatial modelling using a new class of nonstationary
                  covariance functions},
  author =	 {Christopher J. Paciorek and Mark J. Schervish},
  journal =	 {Environmetrics},
  volume =	 17,
  number =	 5,
  pages =	 {483--506, 10.1002/env.785},
  year =	 2006
}

@Article{Paciorek2007,
  title =	 {Bayesian smoothing with {G}aussian processes using
                  {F}ourier basis functions in the {spectralGP}
                  package},
  author =	 {C. J. Paciorek},
  journal =	 {J.~Stat.~Softw.},
  volume =	 19,
  number =	 2,
  pages =	 {nihpa22751},
  year =	 2007
}

@Book{Pawitan2001,
  author =	 {Yudi Pawitan},
  title =	 {In All Likelihood: Statistical Modelling {a}nd
                  Inference Using Likelihood},
  publisher =	 OUP,
  year =	 2001,
  address =	 {New York}
}

@Book{Percival+93,
  author =	 "Donald B. Percival and Andrew T. Walden",
  title =	 "Spectral Analysis {f}or Physical Applications,
                  Multitaper {a}nd Conventional Univariate Techniques",
  publisher =	 CUP,
  year =	 1993,
  address =	 "New York"
}

@Article{Persson2006,
  author =	 {B. N. J. Persson},
  title =	 {Contact mechanics for randomly rough surfaces},
  journal =	 SSR,
  year =	 2006,
  volume =	 61,
  pages =	 {201--227, doi: 10.1016/j.surfrep.2006.04.001}
}

@article{Porcu+2024,
  title =	 {The {M}at{\'e}rn model: {A} journey through statistics,
                  numerical analysis and machine learning},
  author =	 {Emilio Porcu and Moreno Bevilacqua and Robert Schaback and
                  Chris J. Oates},
  journal =	 SSC,
  volume =	 39,
  number =	 3,
  pages =	 {469--492, doi: 10.1214/24-STS923},
  year =	 2024
}

@book{Press+2007,
  author =	 {William H. Press and Saul A. Teukolsky and William
                  T. Vetterling and Brian P. Flannery},
  title =	 {Numerical Recipes: {T}he Art {o}f Scientific Computing},
  publisher =	 CUP,
  address =	 {New York},
  year =	 2007,
  edition =	 3
}

@Article{Rappaport+99,
  author =	 {Nicole J. Rappaport and Alex S. Konopliv and Algis
                  B. Kucinskas and Peter G. Ford},
  title =	 {An improved 360 degree and order model of {V}enus
                  topography},
  journal =	 {Icarus},
  year =	 1999,
  volume =	 139,
  pages =	 {19--31}
}

@Book{Rasmussen+2006,
  author =	 {Carl Edward Rasmussen and Christopher
                  K. I. Williams},
  title =	 {Gaussian Processes for Machine Learning},
  publisher =	 {The MIT~Press},
  year =	 2006,
  address =	 {Cambridge, Mass.}
}

@article{Reed+2002,
  title =	 {Spectral representation of fractional {B}rownian motion in $n$
                  dimensions and its properties},
  author =	 {Irving S. Reed and P. C. Lee and Trieu-Kien Truong},
  journal =	 IEEE-IT,
  volume =	 41,
  number =	 5,
  pages =	 {1439--1451, doi: 10.1109/18.412687},
  year =	 2002
}

@article{Reich+2013,
  title =	 {A nonparametric spatial model for periodontal data with
                  nonrandom missingness},
  author =	 {Brian J. Reich and Dipankar Bandyopadhyay and Howard
                  D. Bondell},
  journal =	 JASTA,
  volume =	 108,
  number =	 503,
  pages =	 {820--831, doi: 10.1080/01621459.2013.795487},
  year =	 2013
}

@article{Rodriguez+74,
  title =	 {The design of rainfall networks in time and space},
  author =	 {Ignacio Rodr{\'\i}guez-Iturbe and Jos{\'e} M. Mej{\'\i}a},
  journal =	 WRR,
  volume =	 10,
  number =	 4,
  pages =	 {713--728, doi: 10.1029/WR010i004p00713},
  year =	 1974
}

@Article{Rosenburg+2011,
  title =	 {Global surface slopes and roughness of the {M}oon
                  from the {L}unar {O}rbiter {L}aser {A}ltimeter},
  author =	 {M. A. Rosenburg and O. Aharonson and J. W. Head and
                  M. A. Kreslavsky and E. Mazarico and G. A. Neumann
                  and D. E. Smith and M. H. Torrence and M. T. Zuber},
  journal =	 JGR,
  volume =	 116,
  pages =	 {E02001, doi: 10.1029/2010JE003716},
  year =	 2011
}

@Article{Sagy+2007,
  author =	 {Amir Sagy and Emily E. Brodsky and Gary J. Axen},
  title =	 {Evolution of fault-surface roughness with slip},
  journal =	 {Geology},
  year =	 2007,
  volume =	 35,
  number =	 3,
  pages =	 {283--286, doi: 10.1130/G23235A.1}
}

@article{Sandwell+2022,
  title =	 {Improved bathymetric prediction using geological information:
                  {SYNBATH}},
  author =	 {David T. Sandwell and John A. Goff and Julie Gevorgian and
                  Hugh Harper and Seung-Sep Kim and Yao Yu and Brook Tozer and
                  Paul Wessel and Walter H. F. Smith},
  journal =	 ESS,
  volume =	 9,
  number =	 2,
  pages =	 {e2021EA002069},
  year =	 2022
}

@book{Severini2001,
  author =	 {Thomas A. Severini},
  title =	 {Likelihood Methods {i}n Statistics},
  year =	 2001,
  publisher =	 OUP,
  address =	 {Oxford, UK}
}

@Article{Sharpton+85,
  author =	 {Virgil L. Sharpton and James W. Head},
  title =	 {Analysis of regional slope characteristics on
                  {V}enus and {E}arth},
  journal =	 JGR,
  year =	 1985,
  volume =	 90,
  number =	 {B5},
  pages =	 {3733--3740, doi: 10.1029/JB090iB05p03733}
}

@Article{Shepard+2001,
  author =	 {Michael K. Shepard and Bruce A. Campbell and
                  Mark. H. Bulmer and Tom G. Farr and Lisa R. Gaddis
                  and Jeffrey J. Plaut},
  title =	 {The roughness of natural terrain: {A} planetary and
                  remote sensing perspective},
  journal =	 JGR,
  year =	 2001,
  volume =	 106,
  number =	 {E12},
  month =	 {32777--32795, doi: 10.1029/2000JE001429}
}

@Article{Simon+2026,
  author =	 {Joel D. Simon and Frederik J. Simons and Jessica C. E. Irving
                  and Wenbo Wu and Masayuki Obayashi and Yong Yu and Yongshun
                  John Chen and Hiroko Sugioka and Yann M. Hello},
  title =	 {Hydroacoustic observations of the 15 {J}anuary 2022 {H}unga
                  {T}onga-{H}unga {H}a'apai eruption: {T}he role of bathymetry
                  along the path},
  year =	 2026,
  journal =	 JGR,
  volume =	 131,
  pages =	 {e2025JB032996, doi: 10.1029/2025JB032996}
}

@Article{Simons+2013,
  author = 	 {Frederik J. Simons and Sofia C. Olhede},
  title = 	 {Maximum-likelihood estimation of lithospheric
                  flexural rigidity, initial-loading fraction and load
                  correlation, under isotropy},
  journal = 	 GJI,
  year = 	 2013,
  volume = 	 193,
  number = 	 3,
  pages =	 {1300--1342, doi: 10.1093/gji/ggt056}
}

@Article{Smith+97,
  author =	 "W. H. F. Smith and David T. Sandwell",
  title =	 "Global sea floor topography from satellite altimetry
                  and ship depth soundings",
  journal =	 "Science",
  year =	 1997,
  volume =	 277,
  number =	 5334,
  pages =	 "1956--1962, doi: 10.1126/science.277.5334.1956"
}

@Article{Song+2003,
  author =	 {Teh-Ru Alex Song and Mark Simons},
  title =	 {Large trench-parallel gravity variations predict
                  seismogenic behavior in subduction zones},
  journal =	 {Nature},
  year =	 2003,
  volume =	 301,
  pages =	 {630--633, doi: 10.1126/science.1085557}
}

@Article{Stein+2004,
  author =	 {Michael L. Stein and Z. Chi and L. J.  Welty},
  title =	 {Approximating likelihoods for large spatial data sets},
  journal =	 JRSSB,
  year =	 2004,
  volume =	 66,
  number =	 2,
  pages =	 {275--296}
}

@Article{Stein95,
  title =	 {Fixed-domain asymptotics for spatial periodograms},
  author =	 {Michael L. Stein},
  journal =	 JASTA,
  volume =	 90,
  number =	 432,
  pages =	 {1277--1288, doi: 10.1080/01621459.1995.10476632},
  year =	 1995
}

@Book{Stein99,
  author =	 {Michael L. Stein},
  title =	 {Interpolation of Spatial Data: Some Theory for
                  Kriging},
  publisher =	 {Springer},
  year =	 1999,
  series =	 {Springer Series in Statistics},
  address =	 {New York}
}

@Article{Stephenson84,
  author =	 "Randell Stephenson",
  title =	 "Flexural models of continental lithosphere based on
                  the long-term erosional decay of topography",
  journal =	 GJRAS,
  year =	 1984,
  volume =	 77,
  pages =	 "385--413"
}

@Article{Sun+2015,
  author =	 {Ying Sun and Kenneth P. Bowman and Marc G. Genton
                  and Ali Tokay},
  title =	 {A {M}at\'ern model of the spatial covariance structure
                  of point rain rates},
  journal =	 SERRA,
  year =	 2015,
  volume =	 29,
  pages =	 {411--416, doi: 10.1007/s00477-014-0923-2}
}

@Article{Sykulski+2019,
  author =	 {Adam M. Sykulski and Sofia C. Olhede and Arthur P. Guillaumin
                  and Jonathan M. Lilly and Jeffrey J. Early},
  title =	 {The debiased {W}hittle likelihood},
  journal =	 {Biometrika},
  volume =	 106,
  pages =	 {251--266, doi: 10.1093/biomet/asy071},
  year =	 2019
}

@Article{Tarantola+84,
  author =	 {Albert Tarantola and Alex Nercessian},
  title =	 {Three-dimensional inversion without blocks},
  journal =	 GJRAS,
  year =	 1984,
  volume =	 76,
  number =	 2,
  pages =	 {299--306, doi: 10.1111/j.1365-246X.1984.tb05047.x}
}

@Article{Valentine+2020,
  title =	 {Global models from sparse data: {A} robust estimate of
                  {E}arth's residual topography spectrum},
  author =	 {Andrew P. Valentine and D Rhodri. Davies},
  journal =	 GC,
  volume =	 21,
  number =	 8,
  pages =	 {e2020GC009240, doi: 10.1029/2020GC009240},
  year =	 2020
}

@Book{Vanmarcke2010,
  author =	 {Erik Vanmarcke},
  title =	 {Random Fields. Analysis and Synthesis},
  publisher =	 {World Scientific},
  year =	 2010,
  address =	 {Singapore},
  edition =	 2
}

@Book{Vanmarcke83,
  author =	 {Erik Vanmarcke},
  title =	 {Random Fields. Analysis and Synthesis},
  publisher =	 {MIT Press},
  year =	 1983,
  address =	 {Cambridge, Mass.},
  edition =	 1
}

@article{Vecchia88,
  title =	 {Estimation and model identification for continuous spatial
                  processes},
  author =	 {Aldo V. Vecchia},
  journal =	 JRSSB,
  volume =	 50,
  number =	 2,
  pages =	 {297--312, doi: 10.1111/j.2517-6161.1988.tb01729.x},
  year =	 1988
}

@Article{Walden+94,
  author =	 "Andrew T. Walden and E. J. McCoy and Donald
                  B. Percival",
  title =	 "The variance of multitaper spectrum estimates for
                  real {G}aussian processes",
  journal =	 IEEE-SP,
  year =	 1994,
  volume =	 2,
  pages =	 "479--482"
}

@article{WangKesen+2023,
  title =	 {Which parameterization of the {M}at{\'e}rn covariance
                  function?},
  author =	 {Kesen Wang and Sameh Abdulah and Ying Sun and Marc G. Genton},
  journal =	 SPS,
  volume =	 58,
  pages =	 {100787, doi: 10.1016/j.spasta.2023.100787},
  year =	 2023
}

@Book{Watson95,
  author =	 {G. N. Watson},
  title =	 {A Treatise {o}n {t}he Theory of {B}essel Functions},
  publisher =	 CUP,
  year =	 1995,
  address =	 {Cambridge, UK},
  edition =	 2
}

@Article{Whittaker+2008,
  author =	 {Joanne M. Whittaker and R. Dietmar M\"uller and
                  Walter R. Roest and Paul Wessel and Walter
                  H. F. Smith},
  title =	 {How supercontinents and superoceans affect seafloor
                  roughness},
  journal =	 {Nature},
  year =	 2008,
  volume =	 456,
  pages =	 {938--941, doi: 10.1038/nature07573}
}

@Article{Whittle53,
  author =	 {Peter Whittle},
  title =	 {Estimation and information in stationary time
                  series},
  year =	 1953,
  journal =	 {Arkiv Mat.},
  volume =	 2,
  number =	 23,
  pages =	 {423--434}
}

@Article{Whittle54,
  author =	 {Peter Whittle},
  title =	 {On stationary processes in the plane},
  volume =	 41,
  number =	 {3--4},
  pages =	 {434--449, doi: 10.2307/2332724},
  year =	 1954,
  journal =	 {Biometrika}
}

@InCollection{Wieczorek2015,
  author =	 {Mark A. Wieczorek},
  title =	 {The gravity and topography of the terrestrial
                  planets},
  booktitle =	 {Treatise on Geophysics},
  pages =	 {153--2193, doi: 10.1016/B978-0-444-53802-4.00169-X},
  publisher =	 {Elsevier},
  year =	 2015,
  editor =	 {Tilman Spohn},
  edition =	 2,
  volume =	 10,
  address =	 {Amsterdam, Neth.}
}

@Article{Wu+85b,
  author =	 {Ru-Shan Wu and Keiiti Aki},
  title =	 {Elastic wave scattering by a random medium and the small-scale
                  inhomogeneities in the lithosphere},
  journal =	 JGR,
  year =	 1985,
  volume =	 90,
  number =	 "B12",
  pages =	 {10261--10273, doi: 10.1029/JB090iB12p10261}
}

@Article{Wu+90,
  author =	 {Ru-Shan Wu and Stanley M. Flatt\'e},
  title =	 {Transmission fluctuations across an array and
                  heterogeneities in the crust and upper mantle},
  journal =	 PAGEOPH,
  year =	 1990,
  volume =	 132,
  number =	 "1--2",
  pages =	 {175--196, doi: 10.1007/BF00874362}
}

@article{Wunsch2022,
  title =	 {Can oceanic flows be heard? {A}byssal melodies},
  author =	 {Carl Wunsch},
  journal =	 JASA,
  volume =	 152,
  number =	 4,
  pages =	 {2160--2168, doi: 10.1121/10.0014603},
  year =	 2022
}

@Article{Zhang+2005b,
  title =	 {Towards reconciling two asymptotic frameworks in spatial
                  statistics},
  author =	 {Hao Zhang and Dale L. Zimmerman},
  journal =	 {Biometrika},
  volume =	 92,
  number =	 4,
  pages =	 {921--936, doi: 10.1093/biomet/92.4.921},
  year =	 2005
}

@Article{Zhang2004,
  title =	 {Inconsistent estimation and asymptotically equal
                  interpolations in model-based geostatistics},
  author =	 {Hao Zhang},
  journal =	 JASTA,
  volume =	 99,
  number =	 465,
  pages =	 {250--261, doi: 10.1198/016214504000000241},
  year =	 2004
}

\vspace{-0.2cm}

\section{A~P~P~E~N~D~I~X}

The explicit forms of eq.~(\ref{mth}), the logarithmic derivatives of the
isotropic Mat\'ern spectral density~(\ref{matern2}) in each of the three
parameters $\theta\in\btheta$, which enter into the definition of the
score~(\ref{score}) and the Fisher matrix~(\ref{fisher}) of the
likelihood~(\ref{lik}), are given in terms of the
auxiliary variable
\begin{equation}
\mu=\left(\frac{4\nu}{\pi^2\rho^2}\right)\left(\frac{4\nu}{\pi^2\rho^2}+k^2\right)^{-1}
,
\end{equation}
by which we rewrite eqs~A25--A27 \cite{Simons+2013} as
\begin{equation}
  m_\st=  \frac{1}{\st},\label{mst}\qquad
  m_\nu=  \frac{\nu+1}{\nu}-\frac{\nu+1}{\nu}\mu+\ln\mu,\also
  m_\rho=-2\frac{\nu}{\rho}+2\frac{\nu+1}{\rho}\mu
.
\end{equation}
The nonvanishing derivatives of~(\ref{mst}) necessary for the computation of the
Hessian~(\ref{hessian}) are
\begin{equation}\label{m1}%\label{m2}\label{m3}
\frac{\pl m_\st}{\pl\st}  =-\frac{1}{\sigma^4},\qquad
\frac{\pl m_\nu}{\pl\nu}  = \frac{1}{\nu}\left(\frac{\nu+1}{\nu}-\mu+\frac{\nu+1}{\nu}\mu^2\right),\qquad
\frac{\pl m_\rho}{\pl\rho}=2\frac{\nu}{\rho^2}\left(1-3\frac{\nu+1}{\nu}\mu+2\frac{\nu+1}{\nu}\mu^2\right),
\end{equation}
and the two identical cross-derivatives are
\begin{equation}\label{m6}
  \frac{\pl m_\nu}{\pl\rho}=\frac{\pl m_\rho}{\pl\nu}=
  \frac{2}{\rho}\left(-1+\frac{2\nu+1}{\nu}\mu-\frac{\nu+1}{\nu}\mu^2\right).
\end{equation}
We list them for completeness only, since they are not needed for our analysis,
though we used them for numerical checks.

Obtaining explicit forms for eq.~(\ref{mthbar}), the logarithmic derivatives of
the blurred spectral density~(\ref{interm}) necessary to calculate the
score~(\ref{bgth}) and the Fisher matrix~(\ref{fisherb}) of the
likelihood~(\ref{blik2}), requires derivatives of the isotropic Mat\'ern spatial
covariance~(\ref{Kr}) in the parameters $\theta\in\btheta$.

The first, in terms of the variance~$\st$, is trivial,
\begin{equation}\label{Cst}\
  \frac{\pl\mcC\ofst\ofsr}{\pl\sigma^2} =
  \frac{2^{1-\nu}}{\Gamma(\nu)}\ofargu^\nu K_\nu\ofargu
  .
\end{equation}

The second, in terms of smoothness~$\nu$, has a compact analytical form when
$\nu=n\in\mathbb{Z}$, for which \cite{Abramowitz+65} eq.~(9.6.45) provide the
integer-order derivative of the modified Bessel function of the second
kind. Using the shorthand $z=\arguni$, we have \cite[see also][]{Geoga+2023},
\begin{equation}\label{Cnu}
  \frac{\pl\mcC\ofst\ofsr}{\pl n} 
  = \sigma^2\frac{2^{1-n}}{\Gamma(n)}z^n
 K_n\ofarguz \left(
\frac{1}{2}+\frac{z}{2}+\ln\left(\frac{z}{2}\right)-\psi(n) 
     +
n!\left(\frac{z}{2}\right)^{-n}
       \sum^{n-1}_{m=0}\frac{\ofarguzh^m K_m\ofarguz}{2(n-m)m!} 
       \right),
       \with \psi(n)=\frac{1}{\Gamma(n)}\frac{d\Gamma(n)}{dn}.
\end{equation}
At large arguments and for non-integer~$\nu$ we resort to a central-difference
approximation that follows the guidance of \cite{Press+2007} eq.~(5.7.8) in
selecting a $\nu$-dependent step size to minimize truncation and round-off
errors, that is $\epsilon^{1/3}\nu$, with $\epsilon$ the machine precision.

The third, with respect to range~$\rho$, is found from the chain rule
and \cite{Watson95} eq.~(3.71.3), $zK_{\nu}'(z)+\nu K_{\nu}(z)=-z K_{\nu-1}(z)$,
\begin{equation}\label{Crho}
    \frac{\pl\mcC\ofst (r)}{\pl\rho} =
    \frac{\sigma^2}{\rho} \frac{2^{1-\nu}}{\Gamma(\nu)} 
    \ofargu^{\nu+1} K_{\nu-1} \left( \argu \right).
\end{equation}

\end{document}